\documentclass[aps,prl,twocolumn,superscriptaddress,amsmath,amssymb,floatfix,nofootinbib]{revtex4-2}
\usepackage{amsmath,amsthm,amsfonts,amssymb,bm,graphicx,color,mathpazo,times, braket}
\usepackage{graphicx}
\usepackage{bm}
\usepackage{siunitx}
\usepackage[colorlinks=true,linkcolor=blue,citecolor=blue,urlcolor=blue]{hyperref}

\usepackage{appendix}
\definecolor{p2col}{rgb}{0.00,0.42,0.20}
\newcounter{SMsection}
\newcounter{SMsubsection}[SMsection]

\renewcommand{\theSMsection}{S\arabic{SMsection}}
\renewcommand{\theSMsubsection}{\Alph{SMsubsection}}

\newcommand{\SMsection}[1]{%
    \refstepcounter{SMsection}%
    \setcounter{SMsubsection}{0}%
    \par\bigskip
    \noindent
    \textbf{\theSMsection.\quad #1}%
    \par\medskip
}

\newcommand{\SMsubsection}[1]{%
    \refstepcounter{SMsubsection}%
    \par\medskip
    \noindent
    \textbf{\theSMsubsection.\quad #1}%
    \par\medskip
}
\makeatother
\begin{document}
\title{Calibration-free quantum thermometry with strain-swept two-level defects}
\author{Asghar Ullah}
 \email{aullah21@ku.edu.tr}
\affiliation{Department of Physics, Ko\c{c} University, 34450 Sar\i yer, Istanbul, T\"urkiye}
\affiliation{Department of Electrical and Electronics Engineering, Ko\c{c} University,
Rumeli Feneri Yolu, Sarıyer, 34450 İstanbul, Türkiye}
\author{\"Ozg\"ur E. M\"ustecapl\i o\u glu}	
	\email{omustecap@ku.edu.tr}
	\affiliation{Department of Physics, Ko\c{c} University, 34450 Sar\i yer, Istanbul, T\"urkiye}
	\affiliation{T\"UBITAK Research Institute for Fundamental Sciences (TBAE), 41470 Gebze, T\"urkiye}
\author{Mehmet Cengiz Onba\c{s}l\i}
\email{monbasli@ku.edu.tr}
\affiliation{Department of Physics, Ko\c{c} University, 34450 Sar\i yer, Istanbul, T\"urkiye}
\affiliation{Department of Electrical and Electronics Engineering, Ko\c{c} University,
Rumeli Feneri Yolu, Sarıyer, 34450 İstanbul, Türkiye}

\date{\today}
\begin{abstract}
Superconducting qubits operate within a well-defined temperature range, making local temperature monitoring important. Here we show that intrinsic two-level defects in Josephson-junction barriers can act as local thermometers by exploiting the shape of their population response, without requiring calibration of the readout offset or contrast. Marginalizing over these unknown parameters imposes an identifiability requirement absent from single-parameter thermometric bounds: for narrow sweeps, usable information scales as the fourth power of the sweep width. Hence, a narrow sweep becomes locally indistinguishable from a change in offset or contrast. We illustrate this principle with a strain-swept two-level defect in a Josephson junction. For a reported defect with tunneling energy $\Delta_{0}/h=\SI{1.3}{\giga\hertz}$, swept along one branch of its hyperbola to \SI{2.69}{\giga\hertz} (an asymmetry excursion of \SI{2.36}{\giga\hertz}, or \SI{8}{\volt} of piezo voltage), the analysis projects a single-defect fit precision of \SI{0.58}{\percent} at \SI{20}{\milli\kelvin}. The resulting response enables temperature estimation without prior calibration of the readout offset and contrast, while the same sweep overdetermines the Boltzmann form and identifies significant deviations from thermality. These results establish intrinsic junction defects as a resource for calibration-free local thermometry and highlight identifiability as a key design criterion for primary temperature sensing.
\end{abstract}

\maketitle
\textit{\textbf{Introduction}}---Superconducting quantum processors are designed around error budgets in which an effective temperature often characterizes residual qubit excitation. Beyond computation, these architectures serve as versatile platforms for high-precision quantum metrology~\cite{Kolosvetov2026}. For the device, however, the relevant quantity is the local temperature in the junction barrier, where microscopic two-level systems (TLSs) contribute to dielectric loss and decoherence~\cite{Anderson01011972,Phillips1972,PhysRevLett.95.210503,Muller_2019,Graaf}.  External primary thermometers, including Coulomb blockade (CBT)~\cite{PhysRevApplied.4.034001,PhysRevLett.73.2903,Bradley2016}, shot-noise thermometry~\cite{praks2026electronicprimarythermometry}, and $^{3}$He melting curve~\cite{Meschke2011}, are located outside the device and do not directly measure this local temperature.
Thermal boundary resistance and electron–phonon decoupling can further produce a mismatch between the measured bath temperature and the local device temperature, with the latter described by $P_{\rm ep}=\Sigma\Omega(T_{e}^{5}-T_{\rm ph}^{5})$~\cite{Bradley2016}. This mismatch is evident in practice: the excited-state population of a transmon has been observed to saturate at an effective floor $T_{\rm eff} \approx \SIrange{30}{45}{\milli\kelvin}$ even when the mixing-chamber bath sits at \SI{15}{\milli\kelvin}~\cite{PhysRevLett.114.240501}. Driven two-level defects can also be pushed into out-of-equilibrium or population-inverted states~\cite{Spiecker2023}.

On-chip secondary thermometers, such as TLS microresonators~\cite{Wheeler2020}, infer temperature from a frequency shift whose amplitude depends on the TLS filling factor and zero-temperature loss tangent. They require calibration against a resistive standard and recalibration of the zero-temperature resonance $f_{0}$ after each cooldown, as frequency shifts can be comparable to the TLS-induced signal~\cite{Wheeler2020}. Similarly, thermometry via single fluctuators in quantum dots~\cite{1fz7-tjq8} or defects inside superconducting barriers~\cite{PhysRevLett.121.090502, PhysRevLett.123.190502, PhysRevLett.105.230504} extract population ratios $e^{-E/k_B T}$, but without an independent measure of the energy $E$, they yield only a temperature ratio instead of an absolute temperature. Qubit-population thermometry avoids amplitude calibration, but at a fixed frequency, it gives only one effective temperature and cannot confirm thermal equilibrium with the bath~\cite{PhysRevLett.124.240501, PhysRevLett.114.240501}. This limitation is evident from the saturation observed below \SI{35}{\milli\kelvin}~\cite{PhysRevLett.114.240501}. Similarly, transmon population thermometry tracks the cryostat temperature only over a limited range of about \SIrange{100}{200}{\milli\kelvin}~\cite{PhysRevApplied.23.054079}. However, qubit coherence depends on microscopic processes within the barrier, making the local temperature near defects particularly relevant to understanding qubit errors. This raises a question: can the intrinsic properties of defects already present in superconducting qubits be exploited to monitor their local operational temperature, turning an otherwise unavoidable source of error into a useful sensing resource?

In this Letter, we show that mechanically tuning a defect's asymmetry energy~\cite{Grigorij, Lisenfeld2019} allows its population to be used for thermometry without calibrating the readout amplitude. The temperature is extracted from the shape of the response, avoiding the amplitude calibration required in conventional TLS thermometry~\cite{Wheeler2020}. We derive the operating point from the quantum Fisher information (QFI): in the low-temperature regime considered here, thermometric sensitivity becomes negligible at the splittings typical of superconducting qubits, so the minimum splitting $\Delta_0$ of the defect must be a few $k_BT$, about \SI{2}{\giga\hertz} or less at \SI{20}{\milli\kelvin}. This exponential degradation of equilibrium thermometry for probes with a lower-bounded gap is known~\cite{Mehboudi_2019} and identifies the low-energy defect regime as the relevant operating window, rather than the larger splittings of standard transmons. Finally, sweeping the energy axis overdetermines the Boltzmann form, so the fit residual reveals significant departures from thermal equilibrium. Such departures are otherwise difficult to detect when the steady state is non-Gibbsian, as when a TLS ensemble is driven through population inversion, where a single-point ratio would still return a finite temperature~\cite{Spiecker2023}.

\textit{\textbf{Hamiltonian and symmetry}}---In the localized double-well basis ${|L\rangle,|R\rangle}$, the system Hamiltonian is
\begin{align}
H = H_{\mathrm{TLS}}+H_{q}+H_{\mathrm{int}}+H_{B}+H_{SB},\label{eq:Htot}
\end{align}
where the Hamiltonians describing the defect, readout qubit, their coupling, and environment are given by~\cite{Muller_2019,RevModPhys.59.1}
\begin{align}
    H_{\mathrm{TLS}} &= \tfrac{1}{2}\bigl[\varepsilon(\epsilon_{s})\,\tau_{z}
                    +\Delta_{0}\,\tau_{x}\bigr],\quad
\varepsilon=\varepsilon_{0}+2\gamma_{S}\epsilon_{s},\label{eq:Htls}\\
H_{q} &= \tfrac{1}{2}\hbar\omega_{q}\Sigma_{z},\qquad
H_{\mathrm{int}} = \hbar g_{0}\,\tau_{z}\otimes\Sigma_{x},\label{eq:Hint}\\
H_{SB} &= \tau_{z}\otimes\hat{B},\qquad
\hat{B}=\textstyle\sum_{k}\lambda_{k}(b_{k}+b_{k}^{\dagger}).\label{eq:Hsb}
\end{align}
Here, \(\epsilon_s=\delta L/L\) denotes the applied strain, \(\gamma_S\) is the deformation potential, \(\varepsilon\) and \(\Delta_0\) denote energies, and \(\omega_q\) and \(g_0\) denote angular frequencies. We follow the superconducting-circuit convention~\cite{Muller_2019}, where $\varepsilon$ and $\Delta_0$ denote the asymmetry and tunneling energy, respectively, unlike the original tunneling-model notation~\cite{Anderson01011972, Phillips1972}. An electric field enters equivalently through $\varepsilon\to\varepsilon+2\mathbf{p}\cdot\mathbf{E}$~\cite{Muller_2019}, but a dc bias applied to a transmon does not generate the corresponding field inside the junction barrier because the potential difference is compensated by Cooper-pair tunneling~\cite{Bilmes2021}. Thus, barrier defects are effectively strain-tuned in the present protocol~\cite{Lisenfeld2019, Bilmes2021}. The qubit coupling, the strain drive, and the defect baths, phonons~\cite{Phillips1972} and photons~\cite{Muller_2019}, all act through $\tau_z$, as they perturb the well asymmetry.
Quasiparticles affect the junction differently: by altering the phase across the junction, they primarily degrade the readout qubit~\cite{Catelani, 1fz7-tjq8, Bogoliubov}.

Diagonalizing Eq.~(\ref{eq:Htls}) with
$\cos\theta=\varepsilon/E$, $\sin\theta=\Delta_{0}/E$, and
$E=(\varepsilon^{2}+\Delta_{0}^{2})^{1/2}$ gives
$H_{\mathrm{TLS}}=E\sigma_{z}/2$. The coupling operator $\tau_{z}$ in terms of $u\equiv\sin^{2}\theta=(\Delta_{0}/E)^{2}$ becomes
\begin{equation}
\tau_{z}=\sqrt{1-u}\,\sigma_{z}-\sqrt{u}\,\sigma_{x}.
\label{eq:rot}
\end{equation}
The parameter $u$ sets the transverse and longitudinal coupling weights. The transverse component drives relaxation,
$\Gamma_{1}=u\,S(E/\hbar)$, while the longitudinal component drives pure dephasing,
$\Gamma_{\phi}=(1-u)\,S(\omega\!\to\!0)$. For a smooth bath, \(\Gamma_1\propto\Delta_0^2/E^2\)~\cite{Lisenfeld2016}, recovering the $\Delta_0^2$ dependence of the one-phonon tunneling model~\cite{Phillips1972}. The defect noise is similarly divided between the frequencies $\pm E$ and zero
frequency, with weights $u$ and $1-u$, respectively~\cite{Muller_2019}. The qubit--TLS swap
element is $g=g_{0}\sqrt{u}$, giving the $\sin\theta$ dependence of the measured avoided-crossing
size~\cite{PhysRevLett.95.210503}.

\begin{figure}[b!]
\centering
\includegraphics[width=\linewidth]{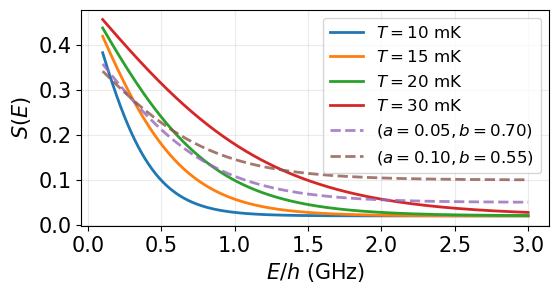}
\caption{TLS response $S(E)$ for several temperatures. Dashed curves show affine transformations of a reference response, illustrating that unknown offset and contrast cannot reproduce the temperature-dependent curvature.
}
    \label{fig:S(E)}
\end{figure}
Three symmetry properties follow, each checked against exact diagonalization.
First, $\tau_{x}\tau_{z}\tau_{x}=-\tau_{z}$ gives
$H_{\mathrm{TLS}}(-\varepsilon)=\tau_{x}H_{\mathrm{TLS}}(\varepsilon)\tau_{x}$,
so the spectrum and all $\tau_z$-coupled rates are even in $\varepsilon$. Thus, $\Gamma_1(\varepsilon)$ is symmetric about the degeneracy point~\cite{Lisenfeld2016}. An odd component in \(\varepsilon\) provides a useful diagnostic. It cannot arise from a bath coupled only through \(\tau_z\), and may indicate a symmetry-breaking mechanism, such as a strain-tuned partner defect. Experimentally, the symmetry of \(\Gamma_1(\varepsilon)\) has been used to rule out near-resonant partner defects as the main cause of relaxation. When a partner defect is tuned through its own symmetry point, the response instead shows an S-shaped curve~\cite{PhysRevB.92.035442, Lisenfeld2015}.
\begin{figure*}[t!]
    \centering
    \includegraphics[width=0.99\linewidth]{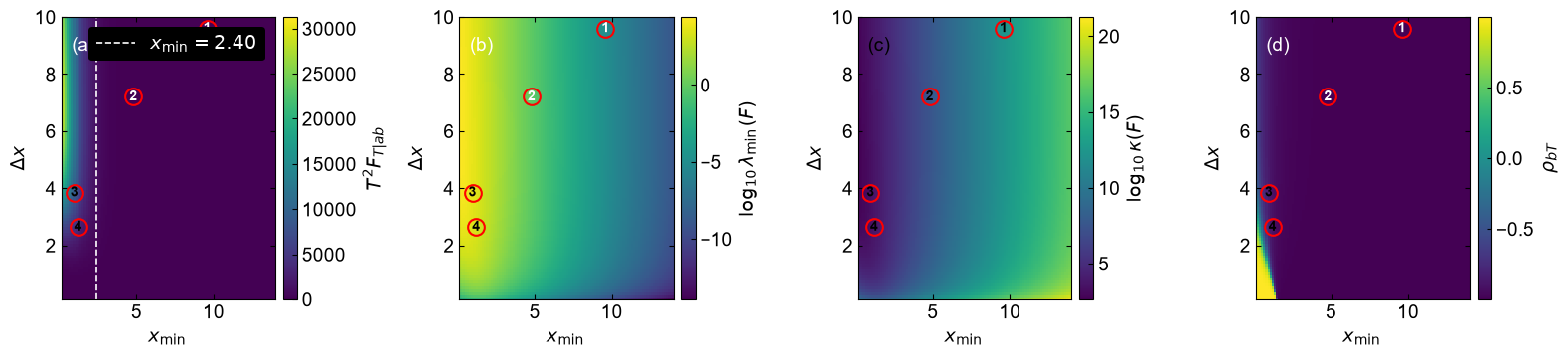}
    \caption{The effective Fisher information $T^{2}F_{T|ab}$, smallest Fisher
eigenvalue $\lambda_{\min}(F)$, condition number $\kappa(F)$, and $b$--$T$ correlation are shown as functions of $x_{\min}=E_{\min}/k_{B}T$ and $\Delta x=(E_{\max}-E_{\min})/k_{B}T$. The marked points indicate the four architecture windows and their frequency ranges: 1, 4--8~GHz (transmon); 2, 2--5~GHz (transmon); 3, 0.5--2.0~GHz (fluxonium); and 4, 0.5--1.6~GHz (low-$f$ probe).}\label{fig:identifiability}
\end{figure*}
Second, at $\varepsilon=0$ one has
$\{H_{\mathrm{TLS}},\tau_{z}\}=0$. The operator that couples longitudinally to the
baths then has strictly vanishing diagonal matrix elements, so first-order pure dephasing is forbidden by symmetry; a quadratic contribution of order $\delta\varepsilon^{2}/2\Delta_0$ remains. Third, within the rotating-wave
approximation the Hamiltonian conserves $\sigma_{z}+\Sigma_{z}$, and in the
one-excitation manifold
\begin{equation}
\lambda_{\pm}=\pm\tfrac{1}{2}\sqrt{\delta^{2}+4\hbar^{2}g_{0}^{2}u},
\qquad
\delta=E-\hbar\omega_{q},
\label{eq:gap}
\end{equation}
so the avoided-crossing gap is $2\hbar g_{0}\sqrt{u}$ which is largest at $u=1$~\cite{SM}. This reproduces the splitting
$S=S_{\max}\cos\eta\sin\theta$ measured in phase-qubit spectroscopy~\cite{PhysRevLett.95.210503},
with $\sin\theta=\sqrt{u}$, and the same $\sqrt{u}$ dressing of the bare coupling
reported in transmon defect spectroscopy,
$g=g'\Delta_{0}/\sqrt{\Delta_{0}^{2}+\varepsilon^{2}}$
~\cite{Lisenfeld2019}. The suppression of dephasing at $\varepsilon=0$ is also reflected in the data: defects with coherence times of several microseconds lie near their symmetry point, whereas typical coherence times are only \SIrange{50}{100}{\nano\second}~\cite{Lisenfeld2019}. Direct strain-sweep measurements similarly show that most defects exhibit strongly suppressed pure dephasing at their degeneracy
points~\cite{Lisenfeld2016}. The degeneracy point is thus simultaneously the point free of first-order dephasing, the point of strongest readout, and, because $\Delta_0>x^{*}k_BT$ for the defect considered here, the point of the sweep closest to the optimum of thermal sensitivity. These three optima coincide, which makes the measurement practical. The coincidence is in $u$ at fixed $g_{0}$; across different defects, coherence can instead be optimal at intermediate coupling~\cite{PhysRevLett.105.177001}, which
is a different parameter axis from that traversed by a single-defect strain sweep.


\textit{\textbf{Principle of the measurement}}---Swap spectroscopy maps the TLS population onto the qubit readout~\cite{PhysRevLett.93.180401,Lisenfeld2015,PhysRevLett.105.177001}. The raw signal is affine in the excited-state occupation
\begin{equation}
S(\varepsilon)=a+bp_{e}\bigl(E(\varepsilon)\bigr),
\qquad
p_{e}=\bigl[1+e^{E/k_{B}T}\bigr]^{-1},
\label{eq:signal}
\end{equation}
where $a$ and $b$ account for readout infidelity, contrast, qubit relaxation during the swap, and the reduced swap visibility associated with the coupling $g_{0}\sqrt{u}$. The key prerequisite is that $a$ and $b$ remain fixed during the sweep while $E$ changes. This is not automatic: a readout qubit tuned across the band acquires a frequency-dependent thermal population, which can contribute a background comparable to the TLS signal unless the qubit and defect share a temperature. We therefore adopt a fixed-frequency readout with a parametric swap, which keeps $a$ independent of $E$~\cite{SM}.

Temperature controls the functional shape of the response $S(E)$. So the hyperbolic strain dependence directly yields $E/h$ in frequency units, while its minimum fixes $\Delta_0$~\cite{Grigorij}. Multiple strain points allow the three unknown parameters $(a,b,T)$ to be extracted simultaneously from the shape of the response, so the temperature inference requires only $h$, $k_{B}$, a frequency scale, and a measured population ratio, without an independently calibrated thermometric standard. As illustrated in Fig.~\ref{fig:S(E)}, changing $T$ uniquely alters the response curvature, whereas variations in offset $a$ or contrast $b$ merely translate or rescale the trace. This extraction of temperature from response shape mirrors the operating principle of normal-metal--insulator--superconductor primary junctions~\cite{PhysRevApplied.4.034001}, but generates its energy axis locally inside the junction barrier. 

However, this parameter identifiability strictly requires sufficient spectral leverage. Narrow or high-energy sweeps diminish the curvature, causing temperature changes to become degenerate with linear offsets and contrast shifts. More generally, for any tunable probe read through an unknown affine readout, the temperature precision in the narrow-sweep regime ($\Delta E \ll k_B T$) and in the readout-floor regime ($a\gg p_e$) obeys
\begin{equation}
\frac{\sigma_T}{T}
\simeq
\frac{\sqrt{720\,a(1-a)}\,e^{E_{\min}/k_B T}}{b\sqrt{N_{\rm tot}}}
\left( \frac{k_B T}{\Delta E} \right)^2.
\label{eq:design_criterion}
\end{equation}
When $p_e$ instead dominates the readout noise, the exponential becomes $e^{E_{\min}/2k_BT}$. The $(\Delta E/k_B T)^{-2}$ scaling, a $\Delta x^{-4}$ dependence of the variance, is the narrow-sweep asymptote ($\Delta x \ll 1$). It does not limit our budget sweep, which has $\Delta x \approx 3.3$ and $p_e(\Delta_0)=4.2\%>a$, and whose $0.58\%$ precision is computed numerically (Sec.~S2C of SM~\cite{SM}). The same structure applies to other platforms: for silicon-germanium charge fluctuators, narrowing the energy window from the accessible \SI{22.4}{\micro\electronvolt} to \SI{2}{\micro\electronvolt} raises the bound from about $2\%$ to above $100\%$ (Sec.~S2A of SM~\cite{SM}).
Thus, sufficient spectral leverage is required not only to obtain high precision, but also to separate temperature from the unknown offset and contrast. Figure~\ref{fig:identifiability} maps this information loss across the accessible sweep parameter space, demonstrating how insufficient spectral leverage leads to severe $b$--$T$ parameter correlation and matrix ill-conditioning. Here, the defect's internal energy is tuned directly within the junction barrier. The residuals of the overdetermined single-temperature fit reveal significant deviations from thermal equilibrium. For a precision-optimized sweep, however, these residuals alone cannot distinguish genuine non-thermality from low-level readout-contrast drift, necessitating the spectroscopic readout protocol detailed below.

\begin{figure*}[t!]
    \centering
    \includegraphics[width=\textwidth]{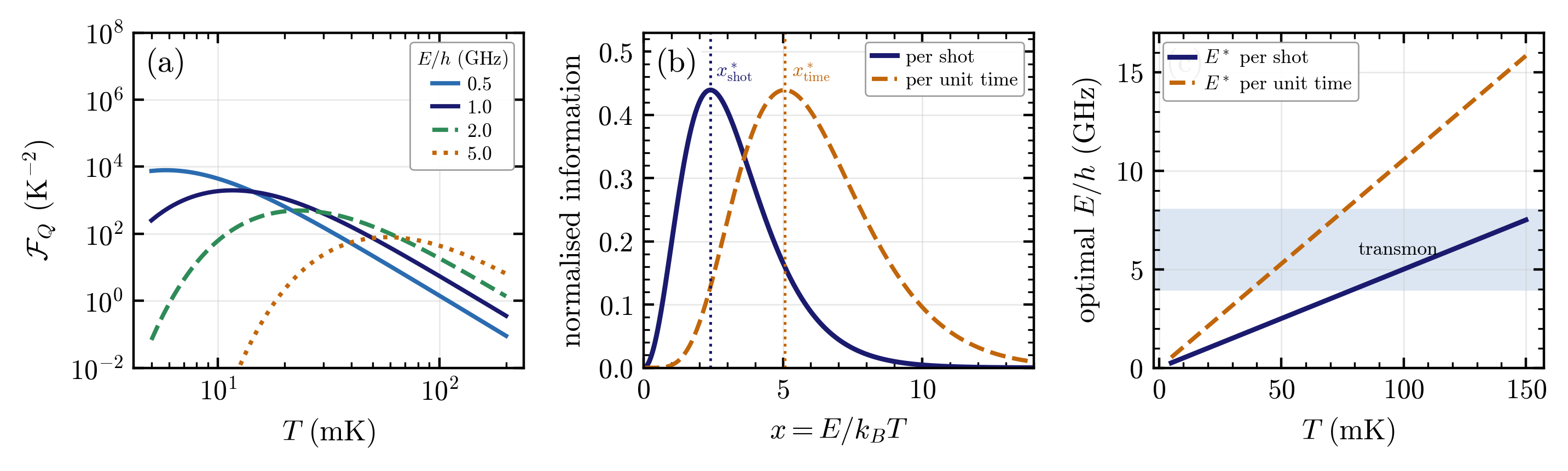}
    \caption{Temperature dependence of the quantum Fisher information.
(a) $\mathcal{F}_{Q}(T)$ for four splittings $E/h$. (b)  Normalized information per shot and per unit time. The
per-shot Fisher information scales as
$x^{2}/\cosh^{2}(x/2)$ and is optimized at $x^{*}=2.40$.
Including the reset cost
$\tau_{\rm shot}\simeq5/\Gamma_{1}$, with
$\Gamma_{1}\propto uE^{3}\coth(x/2)$, the information accumulated
per unit time scales as
$x^{5}\coth(x/2)/\cosh^{2}(x/2)$, shifting the optimum to
$x^{*}\simeq5.0$. At $T=\SI{20}{\milli\kelvin}$ this corresponds to
$E/h\simeq\SI{2.08}{\giga\hertz}$. (c) Optimal splitting versus temperature for both criteria, compared with the transmon band.}
    \label{fig:1}
\end{figure*}

\textit{\textbf{Sensitivity and operating point}}---For a diagonal two-level steady state, the QFI is
\cite{PhysRevLett.114.220405,Mehboudi_2019}
\begin{equation}
\mathcal{F}_{Q}(T)
=\frac{1}{T^{2}}\frac{x^{2}}{4\cosh^{2}(x/2)},
\qquad
 \text{where} \quad x=\frac{E}{k_{B}T}.
\label{eq:qfi}
\end{equation}
The population readout saturates this bound so that the quantum and classical Cram\'er--Rao bounds (QCRB and CRB) coincide. For a transmon probe, the two-level form is an approximation, and retaining
higher anharmonic levels can increase the extractable information~\cite{mandarino2026thermometrymultileveltransmonprobes}. A barrier defect, by contrast, is a genuine double-well TLS with intrawell excitations
at much higher energies, making the two-level description valid in the present
temperature range. That same analysis places the useful transmon window at
\SIrange{50}{500}{\milli\kelvin}, well above our operating point and without the
identifiability constraint.

Equation~(\ref{eq:qfi}) further assumes that the defect equilibrates to the bare Gibbs
state of $H_{\mathrm{TLS}}$. For a system coupled to a bath, the stationary state is the mean-force Gibbs state~\cite{PhysRevLett.127.250601}. Evaluated to second order
in the coupling implied by the measured symmetry-point lifetimes, the resulting bias
in the recovered temperature is below \SI{e-3}{\milli\kelvin}, which is two to three orders of magnitude below the systematic floor, and it increases with the transverse weight $u$~\cite{SM}. The dimensionless quantity $T^{2}\mathcal{F}_{Q}$ peaks at
$x^{*}=2.399$, determined by $e^{x}=(x+2)/(x-2)$~\cite{PhysRevLett.114.220405}, which at
\SI{20}{\milli\kelvin} corresponds to $E/h=\SI{1.00}{\giga\hertz}$
 [Fig.~\ref{fig:1}(a)]. Including the reset cost,
$\tau_{\rm shot}\sim5/\Gamma_{1}$ with $\Gamma_{1}\propto uE^{3}\coth(E/2k_BT)$ for a Debye bath, the information per unit time scales as
$x^{5}\coth(x/2)/\cosh^{2}(x/2)$ and peaks at $x^{*}\simeq5.0$, corresponding to
$E/h=\SI{2.08}{\giga\hertz}$ [Fig.~\ref{fig:1}(b)]. This criterion compares defects at their symmetry points, where $u=1$, and therefore sets the condition $\Delta_0\lesssim5k_BT/h$. Along a single sweep at fixed $\Delta_0$, $u=(\Delta_0/E)^2$ gives $\Gamma_1\propto\Delta_0^{2}E\coth(x/2)$ and moves the per-time optimum to $x\approx3.0$. Optimizing
$\mathcal{F}_{Q}/\tau_{\rm shot}$ is the criterion used for nonequilibrium probes~\cite{PhysRevA.108.062421} and whenever estimation time is scarce ~\cite{ Mehboudi_2019}. Symmetry-point lifetimes of \SIrange{0.4}{6}{\micro\second} have been measured for
barrier defects~\cite{Lisenfeld2016,PhysRevLett.105.177001} with $\Delta_{0}/h$
between about \num{6} and \SI{13}{\giga\hertz}. Rescaling the two bracketing lifetimes, \SI{0.44}{\micro\second} at \SI{6.2}{\giga\hertz} and \SI{3.2}{\micro\second} at \SI{7.34}{\giga\hertz}, with the same $\Gamma_{1}$ scaling gives $T_{1}$ between \SI{23}{\micro\second} and \SI{0.53}{\milli\second} over $E/h=\SIrange{1.3}{2.69}{\giga\hertz}$, so $\tau_{\rm shot}$ lies between \SI{0.11}{\milli\second} and \SI{2.6}{\milli\second}; the maximum thermal-TLS phonon rate
quoted at \SI{20}{\milli\kelvin} is likewise only
$\approx\SI{1.9}{\kilo\hertz}$~\cite{PhysRevLett.123.190502}. The assumed shot duration, $\tau_\text{shot} = \SI{0.30}{\milli\second}$, averaged over the sweep, lies within this window. The reset cost therefore determines the sweep duration, and the optimum $\tau^{*}$ derived below is the time required to acquire the shots in that sweep.
Operating at \SI{5}{\giga\hertz} costs a factor of $496$ in shots, and at \SI{10}{\giga\hertz} a
factor of $2\times10^{7}$. Unlike a pure-dephasing probe, whose disturbance arises from coupling and decoupling~\cite{PhysRevA.108.062421}, the swap removes one quantum from the defect per shot. The reset interval $\tau_{\rm shot}$ therefore limits the rate of measurement-induced disturbance, making the sweep slow by construction.

This temperature-identifiability constraint determines which architecture is viable. Propagating Eq.~(\ref{eq:signal})
through a Fisher analysis in which $a$ and $b$ are marginalized gives temperature uncertainty
$\sigma_{T}=\SI{108}{\milli\kelvin}$ for a \SIrange{4}{8}{\giga\hertz} sweep at $N = 10^6$ shots per point, compared
with \SI{0.048}{\milli\kelvin} for \SIrange{0.4}{2.0}{\giga\hertz}. The failure is one of rank: at $5\,\mathrm{GHz}$ and $20\,\mathrm{mK}$,
$p_e \simeq 6\times10^{-6}$ and is nearly flat across any accessible sweep, so $(a,b,T)$ become degenerate. The same conclusion is reached from
the transmon side: for barrier defects at $E_{\rm TLS}/h\approx\SI{5.5}{\giga\hertz}$
and $T=\SI{15}{\milli\kelvin}$, the Boltzmann factor is estimated at
$\sim10^{-8}$, and the defects are taken to exhibit no thermal dynamics at all
~\cite{PhysRevLett.121.090502}. Primary operation therefore requires the low-frequency band, and the only way to reach it with a readable defect is to strain-tune toward the hyperbola minimum, which is therefore an enabling requirement.
\begin{figure*}[t!]
    \centering
    \includegraphics[width=\textwidth]{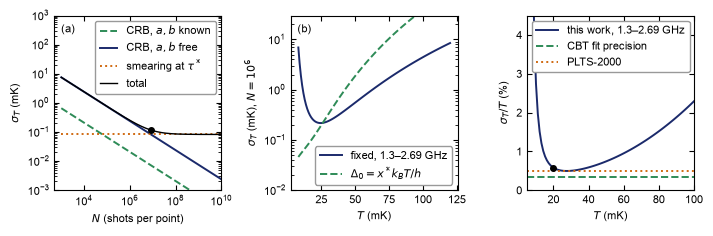}
    \caption{Precision bounds at \SI{20}{\milli\kelvin}.
(a) Lower bound with $a,b$ known, attainable bound with $a,b$ marginalized, the systematic floor, and their quadrature sum. The CRB with $a,b$ known crosses the systematic floor at
$N\simeq1.2\times10^4$, whereas the marginalized CRB crosses it at a total budget of $N_{\text{tot}}^* \approx 9.27 \times 10^7$ shots ($N^* \approx 8.42 \times 10^6$ shots per strain point). The latter defines the crossover relevant to the attainable bound. (b) $\sigma_{T}$ for a fixed sweep and for one that tracks the optimal energy ratio as $T$ varies. (c) Relative uncertainty against CBT and PLTS-2000, reaching $0.58\%$.}
    \label{fig:2}
\end{figure*}
The constraint is the defect-specific manifestation of a general result. For an equilibrium probe
whose gap is bounded below, the temperature uncertainty diverges exponentially as
$T\to0$, with the low-temperature scaling $\sigma_{T}\sim(T^{2}/E_{\min})\,e^{E_{\min}/2k_{B}T}$. Avoiding this exponential suppression requires the relevant energy gap to scale with temperature, such that the dimensionless ratio $E(T)/k_{B}T$ remains finite~\cite{Mehboudi_2019}.
A transmon fixed near \SI{5}{\giga\hertz} is precisely a gap bounded below, and the
tracking curve of Fig.~\ref{fig:2}(b) realizes that escape. What the present analysis adds is the concrete hardware consequence: the failure is a loss of parameter identifiability, and the bound excludes the un-strained standard transmon outright unless the defect's tunneling minimum $\Delta_0$ is properly matched to $k_B T$ and swept over a sufficient strain range.

\textit{\textbf{Precision bounds}}---Figure~\ref{fig:2} collects the precision bounds and
systematic limits. Energy measurements on a Gibbs state give an exponential-family
likelihood, so the single-parameter bound is saturable at finite shot count
~\cite{Mehboudi_2019}. The asymptotic bound strictly applies to the marginalized three-parameter model, so we checked it by Monte Carlo simulation with a binomial maximum-likelihood estimator (MLE) started from randomized values. At $N^{*}$ and \SI{20}{\milli\kelvin} the estimator spread is \SI{0.0817}{\milli\kelvin} against a CRB of \SI{0.0820}{\milli\kelvin}, the bias is \SI{0.001}{\milli\kelvin}, and the empirical coverage is $68.8\%$ at $1\sigma$ and $95.5\%$ at $2\sigma$ [Sec.~S4 of SM~\cite{SM}]. The cooldown provides a reasonably narrow prior range for the temperature, allowing us to use local estimation theory to optimize the measurement around the expected operating point. If the temperature is not known within such a range, a global approach~~\cite{PhysRevLett.127.190402} is needed to avoid bias from the local approximation. The bound with $a$ and $b$ known is nonetheless unattainable in practice, since they are unknown; marginalizing over these nuisance parameters increases uncertainty. Under real-time in-situ spectroscopic tracking, linear first-order frequency drift is continuously compensated. The remaining contribution is the tracking residual, the part of the frequency wander within each dwell that tracking leaves uncorrected, which enters the estimate at first order. We take this residual equal to the untracked intra-dwell wander, the conservative case $\eta=1$ of Sec.~S8 of SM~\cite{SM}, which contributes \SI{0.082}{\milli\kelvin} at the optimal sweep duration. It changes from sweep to sweep and is therefore a random term. The rate is fixed by the measured $D=\SI{2.5}{\mega\hertz}/\sqrt{\mathrm{h}}$, defined through $\sigma(t)=2D\sqrt{t}$ and correspond to an rms wander of \SI{5.0}{\mega\hertz} after one hour~\cite{PhysRevLett.121.090502}. It has since been measured independently for eight defects in different devices, using the same convention. The measured value is $(2.2\pm0.1)\,\si{\mega\hertz}/\sqrt{\mathrm{h}}$~\cite{4ssz-6ctb}. The slow spectral instability of barrier defects has also been reported~\cite{PhysRevLett.123.190502, Burnett2019}. Sub-second adaptive spectroscopy has revealed a faster component of frequency wander, characterized by Ornstein–Uhlenbeck jitter with a correlation time of \SI{32}{\second} and a spectral diffusivity of $D\approx\SI{0.9}{\mega\hertz\squared\per\second}$, alongside telegraphic switching on a timescale of \SI{2.2}{\second}. These dynamics have been observed in independently fabricated devices in two laboratories~\cite{berritta2026adaptivespectroscopyfasttwolevelsystem}. Both are transmon-band measurements which are treated on the same footing as $D$, and are shorter than the \SI{45.8}{\minute} spent at each strain point by factors of $86$ and $1.2\times10^3$. They therefore smear the energy axis and enter only at second order in the fast rms wander $\sigma_f=\SI{5.5}{\mega\hertz}$, contributing \SI{7e-4}{\milli\kelvin}~\cite{SM}. The statistical uncertainty and the tracking residual are equal at $N^{*} \approx 8.42 \times 10^6$ shots per strain point ($9.27 \times 10^7$ in total) [Fig.~\ref{fig:2}(a)], where each contributes \SI{0.082}{\milli\kelvin} at the optimal sweep duration $\tau^* = \SI{7.63}{\hour}$. Combining all contributions in quadrature yields an attainable total uncertainty of $\sigma_T = \SI{0.116}{\milli\kelvin}$, corresponding to a relative precision of $\sigma_T/T = 0.116/20 = 0.58\%$ at $T = \SI{20}{\milli\kelvin}$.  This $0.58\%$ figure represents the projected single-defect fit precision under in-situ tracking, establishing the protocol's attainable floor while leaving inter-defect and cross-device reproducibility to be validated experimentally~\cite{SM}. A probe whose minimum splitting follows $\Delta_0=x^{*}k_BT$ would avoid the exponential loss of a fixed band [dashed curve in Fig.~\ref{fig:2}(b)]; the reported defect cannot do this at \SI{20}{\milli\kelvin}, where its sweep starts at $x=3.12$. For its fixed sweep, Fig.~\ref{fig:2}(c) shows a minimum of $0.50\%$ near \SI{28}{\milli\kelvin} and $0.74\%$ at \SI{50}{\milli\kelvin}; the uncertainty exceeds $1\%$ only below about \SI{14}{\milli\kelvin} and above about \SI{62}{\milli\kelvin}. The budget describes one defect, with $\Delta_0/h = \SI{1.3}{\giga\hertz}$ setting the lower energy scale, and the local slope $\varepsilon/E$ converts frequency wander into temperature uncertainty point by point across the sweep~\cite{SM}. A tracking residual smaller than the untracked wander improves the precision as $\eta^{1/2}$ at the cost of a run longer by $\eta^{-1}$: a single frequency reference per dwell gives $\eta=1/\sqrt{3}$ and $0.44\%$ at $\tau^{*}=\SI{13.1}{\hour}$~\cite{SM}.

\textit{\textbf{Conclusion}}---We proposed that a strain-swept TLS can serve as a candidate primary
thermometer located inside the Josephson barrier. With a projected single-defect fit precision of \SI{0.58}{\percent} (\SI{0.116}{\milli\kelvin}) at \SI{20}{\milli\kelvin} for a reported defect with $\Delta_{0}/h = \SI{1.3}{\giga\hertz}$ swept to \SI{2.69}{\giga\hertz}, useful operation constrains the hyperbola minimum rather than the upper end of the sweep. The sweep must also span enough energy to make the response identifiable, which for this defect requires an asymmetry excursion of \SI{2.36}{\giga\hertz}, or \SI{8}{\volt} along one branch at the measured tuning rate of \SI{295}{\mega\hertz\per\volt} (\SI{16}{\volt} if both branches are swept). Our analysis shows that this requirement is architectural. At \SI{20}{\milli\kelvin}, the transmon band is exponentially penalized, whereas low-frequency probes can reach the sub-percent regime before accounting for the systematic floor. This advantage combines precision with local sensing: the thermometer probes microscopic degrees of freedom inside the barrier that directly contribute to qubit decoherence, allowing local temperature changes to be tracked within the qubit's operating range. The quoted \SI{0.58}{\percent} is the single-defect fit precision under in-situ spectroscopic tracking and excludes the readout-contrast allowance, at most \SI{0.22}{\percent}, of Table~\ref{tab:sm:budget}. Experimental validation should therefore rely on comparison with an independent primary thermometer and on the consistency of the swept response with a single temperature. The residual-based test detects a \SI200 admixture only when it has already biased the inferred temperature by approximately $13\%$ at the operating shot count. The low-frequency TLS temperature affects qubit errors through spectral diffusion, dephasing, and TLS-induced excitation. Applying this temperature to the \SI{5}{\giga\hertz} TLSs that
dominate dielectric loss requires assuming a common local temperature across
the defect population, which our protocol does not test. Three experimental issues remain. First, the tracking residual and the spectral diffusivity in the sub-transmon band set the budget and remain to be measured. Second, the readout-linearity assumption underlying Eq.~(\ref{eq:signal}) requires direct validation, particularly when the defect passes near the readout resonator. Third, the reported defect couples to its readout qubit at only \SI{3}{\mega\hertz}, so the swap readout needs a readout-qubit lifetime of about \SI{12}{\micro\second} or longer. These tests determine whether the projected precision can be achieved in a fully implemented in-barrier thermometer. Thus, the goal is a sufficiently precise thermometer at the location where the dominant qubit errors originate.

\textit{\textbf{Acknowledgments}}--- The authors acknowledge the ERC Proof of Concept grant AURORA (grant no. 101293210).
\bibliographystyle{apsrev4-2}

\bibliography{References.bib}
\begin{figure*}[t!]
\includegraphics[width=\textwidth]{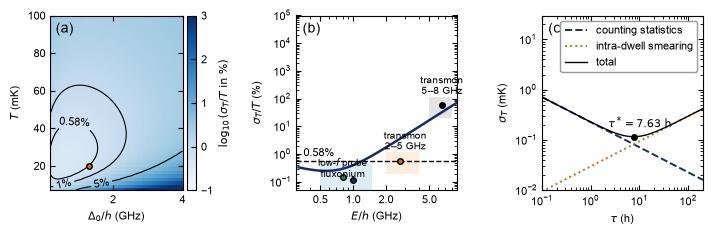}
\caption{(a) Relative uncertainty of the full budget over $(\Delta_{0},T)$ for a sweep that starts at the hyperbola minimum and reaches an asymmetry excursion of \SI{2.36}{\giga\hertz} (\SI{8}{\volt} along one branch); contours mark $0.58\%$, $1\%$ and $5\%$, and the dot marks the reported defect at \SI{20}{\milli\kelvin}. (b) Statistics-limited $\sigma_T/T$ at $N=10^{6}$ shots per point for five sweep windows: transmon, 4--8 and \SIrange{2}{5}{\giga\hertz}; fluxonium, \SIrange{0.5}{2.0}{\giga\hertz}; a dedicated low-frequency probe, \SIrange{0.5}{1.6}{\giga\hertz}; and the reported defect, \SIrange{1.3}{2.69}{\giga\hertz}. (c) Sweep-duration optimum $\tau^* = \SI{7.63}{\hour}$ set by the competition between counting statistics ($\propto \tau^{-1/2}$) and the intra-dwell tracking residual ($\propto \tau^{1/2}$), whose variance grows as $\tau/(N_{\mathrm{pts}}-1)$.}
\label{fig:3}
\end{figure*}
\begin{widetext}
\section*{End Matter}
\end{widetext}
\textit{\textbf{Predicted configurations}}---Figure~\ref{fig:3}(a) maps the achievable relative uncertainty over tunneling energy and temperature, with the black contour marking $0.58\%$. At \SI{20}{\milli\kelvin} and $N=10^{6}$ shots per point, before the tracking residual is included, the statistics-limited $\sigma_T/T$ is $54\%$ for a transmon swept over \SIrange{4}{8}{\giga\hertz}, $4.4\%$ over \SIrange{2}{5}{\giga\hertz}, $0.28\%$ for a fluxonium over \SIrange{0.5}{2.0}{\giga\hertz}, $0.39\%$ for a dedicated low-frequency probe over \SIrange{0.5}{1.6}{\giga\hertz}, and $1.2\%$ for the reported defect over \SIrange{1.3}{2.69}{\giga\hertz}~[Fig.~\ref{fig:3}(b)]. The reported defect is 0.58\% as its shorter reset time allows more shots in the same run. These separate the architectures by several orders of magnitude: only probes operating in the sub-transmon band can reach the sub-percent statistics-limited regime, selecting fluxonium~\cite{PhysRevLett.113.247001} or a purpose-built low-frequency probe. TLS-resonator thermometry faces a similar frequency constraint, with devices operating at \(\SI{2}{\giga\hertz}\) usable only above \(\SI{50}{\milli\kelvin}\)~\cite{Wheeler2020}. The band is not hypothetical: a granular-aluminum fluxonium at $f_{01}=\SI{1.2}{\giga\hertz}$ lies right at the \SI{1.3}{\giga\hertz} tunneling minimum scale of our recommended defect and is already coupled to a TLS environment~\cite{Spiecker2023}. There, the defects are weakly and collectively coupled ($g\gtrsim2\pi\times\SI12$, with no resolved avoided crossings), so the device does not meet the additional requirement for the single-defect swap readout described by Eq.~\eqref{eq:gap}.

\textit{\textbf{Comparison with other in situ approaches}}---Nho \emph{et al.}
~\cite{ql6q-wfpn} obtain
$k_BT_q=hf_q/\ln(\Gamma_{10}/\Gamma_{01})$ from the two transition rates of a
gap-engineered, offset-charge-sensitive 3D transmon, and resolve the substrate
returning from \SI{90}{\milli\kelvin} after a quasiparticle burst
~\cite{Vepsalainen2020,PRXQuantum.4.020356,McEwen2022} with a \SI{6}{\milli\second} time constant. The regimes are complementary. Their qubit frequencies, $f_q=\SIrange{3.57}{4.78}{\giga\hertz}$, are optimal according to Eq.~\eqref{eq:qfi} at \SI{90}{\milli\kelvin}, but lie 9--50 times away from the optimum at their \SI{25}{\milli\kelvin} base temperature. Their fixed-frequency population ratio resolves millisecond-scale dynamics, whereas our method uses a static frequency sweep. More importantly, a single population ratio gives a temperature estimate but cannot establish whether the populations follow a Boltzmann distribution. Our sweep tests this by fitting the Boltzmann form across multiple frequencies, with the fit residual revealing deviations from thermal equilibrium. The source explicitly states that its steady-state value of approximately $\SI{50}{\milli\kelvin}$ does not represent the sample temperature, being
dominated by measurement error~\cite{ql6q-wfpn}. A fast secondary probe and a slow primary one answer different questions; the natural use is to calibrate the former against
the latter. The same holds for the co-fabricated Dayem-bridge thermometer~\cite{lachman2026localizedthermometrydayembridges}, which reads the chip temperature from the critical current without microwave calibration but remains secondary because of the material dependence of $I_c(T)$.
\begin{table}[h]
\caption{\label{tab:cmp}Comparison with sub-kelvin thermometry. ``In situ'' denotes
location inside the device under test; $\tau_{\rm res}$ is the temporal resolution.
Quoted uncertainties are those reported at or near \SI{20}{\milli\kelvin} where
available; note that the \SI{1}{\percent} figure often quoted for CBT applies above
\SI{250}{\milli\kelvin}~\cite{Meschke2011}.}
\begin{ruledtabular}
\begin{tabular}{lccccc}
Method & Type & $\sigma_{T}/T$ & In situ & Wiring & $\tau_{\rm res}$\\
\colrule
CBT~\cite{PhysRevLett.73.2903,Meschke2011} & primary & $0.35\%$\footnotemark[1] & no & dc & s\\
Shot noise~\cite{praks2026electronicprimarythermometry} & primary & $\pm15\%$\footnotemark[2] & no & rf & s\\
N-I-S junction~\cite{PhysRevApplied.4.034001} & primary & 6--11\% & chip & dc & s\\
$^{3}$He curve~\cite{Meschke2011} & primary & $0.5\%$ & no & cell & s\\
RuO$_{2}$/Cernox~\cite{Yeager2002} & secondary & $\gtrsim10\%$\footnotemark[3] & no & dc & ms\\
TLS resonator~\cite{Wheeler2020} & secondary & calib.\footnotemark[4] & chip & none & ms\\
Qubit ratio~\cite{ql6q-wfpn} & secondary & $\sim10\%$ & chip & none & ms\\
Qubit population~\cite{PhysRevApplied.23.054079} & secondary & $\sim9\%$\footnotemark[5] & chip & none & s\\
\colrule
This work & candidate primary & $0.58\%$\footnotemark[6] & barrier & piezo & h\\
\end{tabular}
\end{ruledtabular}
\footnotetext[1]{Fit precision~\cite{praks2026electronicprimarythermometry}; CBT and shot-noise thermometers
inter-compare to \SI{2.5}{\percent} over \SIrange{20}{235}{\milli\kelvin}.}
\footnotetext[2]{At \SI{20}{\milli\kelvin}, dominated by measurement-bandwidth
uncertainty; sub-percent at higher temperature.}
\footnotetext[3]{Operational bound derived from thermal resistance, not factory calibration.}
\footnotetext[4]{No entry at \SI{20}{\milli\kelvin}: the demonstrated range is
\SIrange{50}{1000}{\milli\kelvin}, calibrated against RuO$_{2}$, with
$2\sigma_{T}\simeq\SI{2}{\milli\kelvin}$ over \SIrange{50}{250}{\milli\kelvin}.}
\footnotetext[5]{Three-level population readout; the demonstrated working range is
\SIrange{60}{200}{\milli\kelvin}, the effective qubit temperature saturates at
\SIrange{65}{85}{\milli\kelvin} at a \SI{22}{\milli\kelvin} base, and it tracks the
cryostat only above about \SI{100}{\milli\kelvin}~\cite{PhysRevApplied.23.054079}.}
\footnotetext[6]{Projected single-defect fit precision at \SI{20}{\milli\kelvin} under in-situ spectroscopic tracking; excludes the readout-contrast allowance of Table~\ref{tab:sm:budget}.}
\end{table}

\textit{\textbf{Remaining assumptions}}---Three assumptions remain. First, strain is assumed to leave $\Delta_0$ unchanged. Although strain-induced changes in the tunneling amplitude are generally considered negligible~\cite{Lisenfeld2015}, anticorrelated fluctuations of $\varepsilon$ and $\Delta_0$ have been proposed to explain the coupling dependence of TLS coherence~\cite{PhysRevLett.105.177001}. The residuals from the hyperbola fit can reveal deviations from this assumption.
\begin{figure}[b!]
    \centering
    \includegraphics[width=\linewidth]{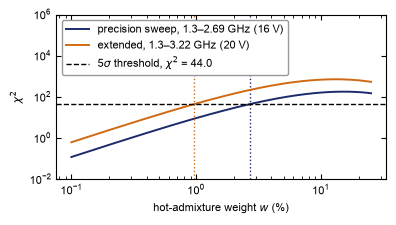}
    \caption{Residual $\chi^2$ versus the weight $w$ of a \SI{200}{\milli\kelvin} hot admixture at $N^{*}$ shots per point, for the precision sweep (\SIrange{1.3}{2.69}{\giga\hertz}) and an extended sweep (\SIrange{1.3}{3.22}{\giga\hertz}). The dashed line marks the $5\sigma$ threshold ($\chi^2 = 44.0$); the dotted lines mark $w_{\min} \approx 2.7\%$ and $0.97\%$.}
    \label{fig:residual}
\end{figure}
Second, a defect remains trackable over a multi-hour sweep. Telegraphic
switching~\cite{PhysRevB.92.035442,PhysRevLett.123.190502} occurs on $\tau_{\rm sw}=\SI{2.2}{\second}$ and diffusive jitter on $\tau_c=\SI{32}{\second}$~\cite{berritta2026adaptivespectroscopyfasttwolevelsystem}. Both were measured in the transmon band, as was $D$, and both are much faster than the hours-scale jump observed for the same reference. Since both timescales are short compared with the \SI{45.8}{\minute} per strain point, their effects average out within each point and bias the fit only through the curvature of $p_e$~\cite{SM}. Relative to the \SI{0.30}{\milli\second} duty cycle, the switching is resolved shot by shot, because a \SI{100}{\mega\hertz} excursion leaves the swap band. The data can therefore be separated by state, with each branch fitted independently to obtain two temperatures whose agreement provides an additional test of thermality. Without state selection, a $\pm\SI{50}{\mega\hertz}$ excursion would increase the total uncertainty from $0.58\%$ to $0.65\%$~\cite{SM}. A slow shift during the run can bias individual measurements without changing the overall estimate. The fit residual may reveal the shift, but cannot distinguish it from instrumental distortions over the operating band~\cite{SM}. The residual quantifies this vulnerability as shown in Fig.~\ref{fig:residual}. At $N^{*}$ shots per point, the precision sweep detects a \SI{200}{\milli\kelvin} hot admixture of weight $w_{\min}\approx2.7\%$ at the $5\sigma$ threshold ($\chi^2=44.0$). Extending the sweep to \SI{3.22}{\giga\hertz}, requiring an asymmetry excursion of \SI{2.95}{\giga\hertz} (\SI{10}{\volt} along one branch), lowers this to $0.97\%$. Detection comes late: at $w_{\min}$, the fitted temperature is already biased by \SI{2.6}{\milli\kelvin} ($13\%$) for the precision sweep and by \SI{1.3}{\milli\kelvin} for the extended sweep. An admixture of just $0.12\%$ biases the temperature by $1\sigma$, while $\chi^2\approx0.17$. The residual therefore reveals large deviations from thermality but cannot confirm thermality at the quoted precision. Moreover, the residual shape of a hot admixture is strongly correlated with that of readout-contrast drift over both sweeps ($|\cos\theta|=0.97$ and $0.93$), so the residual detects a deviation without identifying its cause~\cite{SM}. In an 82-defect survey, about $5\%$ shifted spontaneously; the rest remained stable~\cite{PhysRevLett.105.177001}. Third, and most restrictively, is that a suitable defect can be found. Transmon defect spectroscopy finds a junction-barrier density of $15~\si{\giga\hertz}^{-1}$ per qubit, but only $3\%$ of detected defects show avoided crossings above \SI{5}{\mega\hertz}~\cite{Lisenfeld2019}. Including the stability fraction and the log-uniform $\Delta_0$ distribution gives $0.03$–$0.05$ qualifying defects per qubit. A 90\% chance of finding at least one usable defect therefore requires about $45$–$70$ junctions, which can be accommodated on a single multi-qubit chip. Junction area is also important: in-barrier defects couple to their qubit at a few to tens of megahertz (\SI{3}{\mega\hertz} for the reported defect), favoring the larger junctions used in strain-tuning experiments~\cite{SM}.

The main restriction is that reported defects swept through their symmetry point have
$\Delta_0/h$ between \SI{6}{\giga\hertz} and \SI{13}{\giga\hertz}~\cite{Lisenfeld2015,PhysRevLett.105.177001},
well above the required band. This apparent clustering is partly a selection effect,
since the symmetry point is resolvable only when it lies within the available tuning
range~\cite{Grigorij}. A barrier defect with
$\Delta_0/h=\SI{1.3}{\giga\hertz}$ has been strain-tuned through its symmetry point, which lies at a piezo voltage of \SI{-14.05}{\volt}, demonstrating that low-$\Delta_0$ defects exist~\cite{Lisenfeld2015}. At
\SI{20}{\milli\kelvin} it gives $p_e=4\%$, compared with $6\times10^{-6}$ at
\SI{5}{\giga\hertz}. Its asymmetry tunes at \SI{295}{\mega\hertz\per\volt}, so the sweep to \SI{2.69}{\giga\hertz} needs \SI{8}{\volt} along one branch, from \SI{-14.05}{\volt} to about \SI{-6}{\volt}, a range covered by the reported measurements. A symmetric two-branch sweep would need $-22\text{ V}$ to $-6\text{ V}$, within the $-25\text{ V}$ to $+7\text{ V}$ span of the reported measurements  (Sec.~S11 of SM~\cite{SM}). The defect was never observed directly. It couples to the qubit at only \SI{3.0}{\mega\hertz}, and its parameters were inferred through a strongly coupled partner defect in the qubit band (longitudinal coupling \SI{-428}{\mega\hertz}); its $T_1$ was not measured~\cite{Lisenfeld2015}. At \SI{20}{\milli\kelvin}, the partner, near \SI{6}{\giga\hertz}, remains in its ground state (thermal population $\approx e^{-15}$) and only shifts the hyperbola statically, which the fit absorbs. The weak coupling increases the readout-qubit lifetime required to meet the contrast allowance from \SI{0.58}{\micro\second} at \SI{60}{\mega\hertz} to about \SI{12}{\micro\second} (Sec.~S12 of SM~\cite{SM}). Direct swap readout of this defect at its symmetry point remains the key experimental step.

\clearpage
\onecolumngrid

\begin{center}
\textbf{\large Supplemental Material:}\\[2pt]
\textbf{\large Calibration-free quantum thermometry with strain-swept two-level defects}
\end{center}
\begin{center}

Asghar Ullah$^{1}$, 
\"Ozg\"ur E. M\"ustecapl\i o\u glu$^{1,2}$, 
Mehmet Cengiz Onba\c{s}l\i$^{1,3}$

\vspace{5pt}

\textit{
$^{1}$Department of Physics, Ko\c{c} University, 
34450 Sar\i yer, Istanbul, T\"urkiye
}

\textit{
$^{2}$T\"UBITAK Research Institute for Fundamental Sciences, 
41470 Gebze, T\"urkiye
}

\textit{
$^{3}$Department of Electrical and Electronics Engineering, 
Ko\c{c} University, Rumeli Feneri Yolu, 
Sar\i yer, 34450 Istanbul, T\"urkiye
}
\vspace{5pt}
\textit{(Dated: \today)}
\end{center}

\setcounter{section}{0}
\renewcommand{\thesection}{S\arabic{section}}
\renewcommand{\thesubsection}{\Alph{subsection}}
\renewcommand{\thesubsubsection}{\arabic{subsubsection}}

\setcounter{equation}{0}
\renewcommand{\theequation}{S\arabic{equation}}

\setcounter{figure}{0}
\renewcommand{\thefigure}{S\arabic{figure}}

\setcounter{table}{0}
\renewcommand{\thetable}{S\arabic{table}}
\SMsection{Units and conventions}\label{Units+Conv}

Throughout, we quote energies as frequencies, $E \equiv E/h$ in \si{\giga\hertz}, and
use $k_{B}/h = \SI{20.8366}{\giga\hertz\per\kelvin}$, so that
\begin{equation}
x \equiv \frac{E}{k_{B}T}
= \frac{E[\si{\giga\hertz}]}{\SI{20.8366}{\giga\hertz\per\kelvin}\times T[\si{\kelvin}]}.
\end{equation}
At the reference temperature $T_{0}=\SI{20}{\milli\kelvin}$ this gives
$k_{B}T_{0}/h = \SI{0.41673}{\giga\hertz}$ and $k_{B}T_{0}/\hbar =
\SI{2.6184e9}{\radian\per\second}$. We henceforth use scaled units in which all frequencies are measured in units of $k_B T/\hbar$, such that $\beta=1$ and the operating point is $x^{*}=2.3994$.

The two-level Hamiltonian, the internal coordinate $u$, and the rotation to the
energy eigenbasis are as in the main text,
\begin{equation}
\tau_{z}=\sqrt{1-u}\,\sigma_{z}-\sqrt{u}\,\sigma_{x},
\qquad u=\left(\Delta_{0}/E\right)^{2},
\label{eq:sm:rot}
\end{equation}
with $E=\sqrt{\varepsilon^{2}+\Delta_{0}^{2}}$.

\SMsection{Estimator and Fisher information analysis}
\label{sec:est}
The raw signal is defined as
\begin{equation}
    S(\varepsilon)=a+b\,p_e(E(\varepsilon)),
\end{equation}
where
\begin{equation}
    p_e(E)=\frac{1}{1+e^{E/k_BT}}.
\end{equation}
The excited-state occupation probability of the TLS and the signal parameters $a$ and $b$ are unknown. For $N$ binomial shots at each of a set of splittings $\{E_i\}$, the three-parameter Fisher information matrix for $(a,b,T)$ is
\begin{equation}
\begin{split}
F&=\sum_i
\frac{N}{\mu_i(1-\mu_i)}
\,\bm{g}_i\bm{g}_i^{\mathsf T},
\end{split}
\label{eq:sm:fisher}
\end{equation}
where
\begin{equation}
    \bm{g}_i
=\nabla_{(a,b,T)}\mu_i
=
\left(
1,\;
p_{e,i},\;
b\,p_{e,i}(1-p_{e,i})\frac{x_i}{T}
\right),
\end{equation}
and
\begin{equation}
    \mu_i=a+b\,p_{e,i},
    \qquad
    x_i=\frac{E_i}{k_BT}.
\end{equation}
After eliminating the nuisance parameters $(a,b)$, the effective
Fisher information for temperature is given by the Schur complement
\begin{equation}\label{Schur_complement}
    F_{T|ab}
    =
    \bm q^{\mathsf T}W
    \left[
    I-P(P^{\mathsf T}WP)^{-1}P^{\mathsf T}W
    \right]\bm q,
\end{equation}
where
\begin{equation}
    P_i=(1,p_{e,i}),
    \qquad
    q_i=b\,p_{e,i}(1-p_{e,i})\frac{x_i}{T},
    \qquad
    W=\operatorname{diag}(w_i),
\end{equation}
with
\begin{equation}
    w_i=\frac{N}{\mu_i(1-\mu_i)}.
\end{equation}
Thus, temperature is estimable only from the component of the
temperature response that is $W$-orthogonal to both an additive offset
and a rescaling of the population curve. The corresponding
single parameter CRB for $m=1$ repetitions is $\sigma_T=\sqrt{(F^{-1})_{TT}}$.
Unless stated otherwise, we use $a=0.02$, $b=0.95$, $11$ strain points,
and $N=10^6$ shots per point. 

\SMsubsection{Application to semiconductor quantum-dot charge fluctuators}
\label{sec:sm_qd_tlf}
To show that the Schur-reduced criterion in Eq.~\eqref{Schur_complement} applies outside superconducting architectures, we evaluate it for silicon-germanium two-level charge fluctuators~\cite{1fz7-tjq8}. In such devices, pulse heating is inferred from changes in occupation bias, such that
\begin{equation}
    B = e^{-\Delta E / k_B T}.
\end{equation}
However, uncalibrated charge-sensing offset $a$ and contrast $b$ complicate absolute thermometry, forcing studies restricted to fixed or narrow tuning ranges to report relative temperature ratios $R = T/T_0$. The mathematical cause is captured by Eq.~\eqref{Schur_complement}: in the narrow-sweep regime ($\Delta E \ll k_B T$), projecting out $(a,b)$ causes the effective Fisher information to collapse as $F_{T|ab} \propto (\Delta x)^4$, where $\Delta x = \Delta E / k_B T$ and drive the relative uncertainty as $\sigma_T / T \propto (\Delta x)^{-2}$.

Using gate-lever-arm spectroscopy ($\lambda_P \approx \SI{5.6}{\micro\electronvolt\per\milli\volt}$), an asymmetry tuning range of $\Delta E_{\mathrm{sweep}} \approx 22.4\,\mu\mathrm{eV}$ ($\Delta E / h \approx 5.42\,\mathrm{GHz}$) is experimentally accessible. At $T=40\,\mathrm{mK}$ ($k_BT \approx 3.45\,\mu\mathrm{eV}$), this corresponds to a wide sweep span of $\Delta E / k_BT \approx 6.5$. For an illustrative single-shot charge-sensing benchmark ($a=0.05, b=0.90, T=40\,\mathrm{mK}$, $M=25$ points, $N=3600$ shots/point)~\cite{1fz7-tjq8}, exact evaluation of $F_{T|ab}$ for a window starting at the symmetry point yields a statistical lower bound of $\sigma_T / T = 1.8\%$. Restricting the window to \SIrange{0}{2.0}{\micro\electronvolt} ($\Delta E/h < 0.48\,\mathrm{GHz}$) lowers $F_{T|ab}$ by about four orders of magnitude and gives $\sigma_T/T\approx190\%$. The same \SI{2}{\micro\electronvolt} window started at \SI{2}{\micro\electronvolt} gives $80\%$, so the narrow-window penalty depends on where the window sits. The published measurement already spans the wide window, so the narrow-window figures illustrate the criterion and do not describe that experiment. Equation~\eqref{Schur_complement} thus sets the spectral window a calibration-free measurement needs on this platform.
\SMsubsection{Readout floor and parameter identifiability}

Two distinct mechanisms limit the temperature information at large splitting. First, when the background dominates the excited-state population, $a\gg p_e$, the binomial variance approaches $\mu(1-\mu)\simeq a(1-a)$. Even when $a$ and $b$ are known, the Fisher information for temperature $T$ then scales as
\begin{equation}
    F_{TT}\propto e^{-2x},
\end{equation}
reflecting the readout floor set by the background level. Second, when $a$ and $b$ are estimated jointly with $T$, marginalising $(a,b)$ introduces an additional identifiability penalty,
with the corresponding information vanishing as $(\Delta x)^4$ for
narrow sweeps. This loss is distinct from the readout floor:
it arises because temperature changes become increasingly
indistinguishable from an offset and a rescaling of the population
curve. The resulting loss of parameter identifiability is the key
limitation addressed here.
\SMsubsection{Asymptotic behavior}
\begin{figure}[t!]
    \centering
    \includegraphics[width=0.7\linewidth]{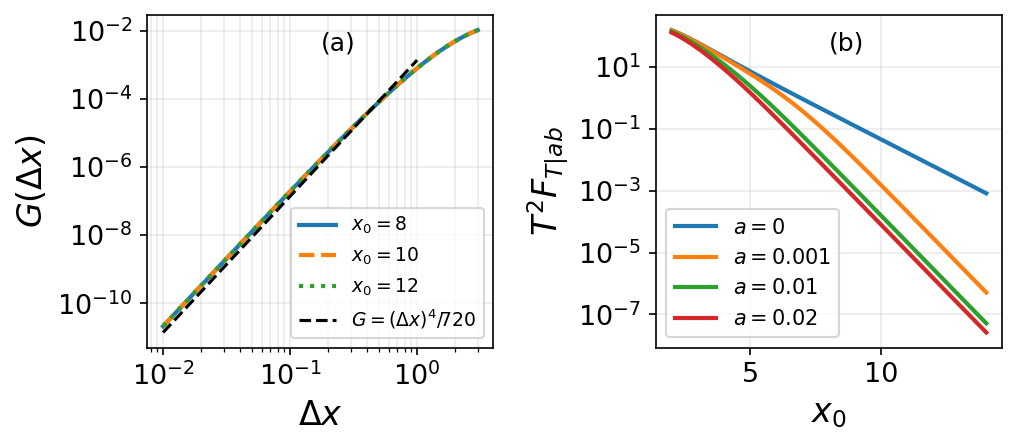}
    \caption{Numerical verification of the asymptotic Fisher-information
scaling. (a) The shape factor $G(\Delta x)$ versus sweep width
$\Delta x$ for several values of $x_0$, together with the narrow-sweep
prediction $G(\Delta x)=(\Delta x)^4/720$. (b) The effective temperature
information $T^2F_{T|ab}$ versus $x_0$ for different readout offsets
$a$, illustrating the crossover between the $e^{-x_0}$ and
$e^{-2x_0}$ regimes.}
    \label{fig:Verifcation_plot}
\end{figure}
To expose the origin of the large-gap suppression, we write
$x=x_0+s$. In the finite-floor regime, where
$\mu\simeq a$, the temperature sensitivity is proportional to
$s e^{-s}$ after factoring out the overall scale $e^{-x_0}$.
Since $x_0e^{-s}$ lies in the nuisance subspace spanned by
$\{1,e^{-s}\}$, the offset contribution is removed by the
Schur-complement projection, giving
\begin{equation}
    F_{T|ab}
    =
    \frac{b^2N_{\rm tot}}
    {a(1-a)T^2}
    e^{-2x_0}
    \left\|
    \Pi_{\{1,e^{-s}\}^{\perp}}
    \left[s e^{-s}\right]
    \right\|^2,
\end{equation}
where the projection is taken with respect to the shot-weighted
inner product. Thus, the remaining shape factor depends only on the
sweep width $\Delta x$ and shot distribution. For a narrow sweep, let $s\in[0,L]$ with $L=\Delta x$. Using
$1-e^{-s}\in\operatorname{span}\{1,e^{-s}\}$, one obtains
\begin{equation}
    s e^{-s}-
    \left(1-e^{-s}\right)
    =
    -\frac{s^2}{2}+O(s^3).
\end{equation}
The mean-square residual of $s^2$ after projection onto
$\operatorname{span}\{1,s\}$ over $[0,L]$ is $L^4/180$, yielding
\begin{equation}
    G(\Delta x)
    =
    \frac{(\Delta x)^4}{720}
    +O\!\left((\Delta x)^6\right),
\end{equation}
and hence
\begin{equation}
    \frac{\sigma_T}{T}
    \simeq
    \frac{\sqrt{720\,a(1-a)}\,e^{x_0}}
    {b\sqrt{N_{\rm tot}}\,(\Delta x)^2}.
\end{equation}
We validate these asymptotic scaling behaviors in Fig.~\ref{fig:Verifcation_plot}. For narrow energy sweeps ($\Delta x \ll 1$), the normalized temperature information collapses onto the universal scaling curve $G(\Delta x) = (\Delta x)^4 / 720$ across different base frequencies $x_0$, which coincide identically in Fig.~\ref{fig:Verifcation_plot}(a). The $\Delta x^4$ scaling law represents the narrow-sweep asymptote and serves as an analytical benchmark; the full $0.58\%$ figure is evaluated numerically across the entire sweep ($p_e = 4.2\%\text{--}0.16\%$, $\Delta x \approx 3.3$). Furthermore, evaluating $T^2 F_{T\vert{}ab}$ across non-zero readout offsets $a$ confirms a clear scaling crossover [Fig.~\ref{fig:Verifcation_plot}(b)]: while an ideal readout ($a = 0$) decays as $\propto e^{-x_0}$, any finite readout floor $a > 0$ accelerates information loss to $\propto e^{-2x_0}$ once $x_0 \gtrsim -\ln a$.

\begin{figure}[t]
    \centering
    \includegraphics[width=0.46\textwidth]{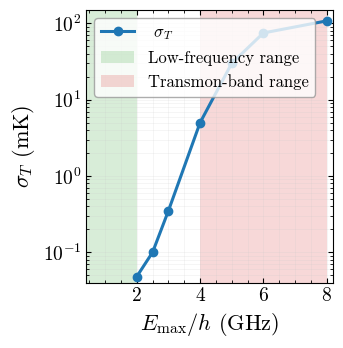}
    \caption{Temperature uncertainty $\sigma_T$ as a function of the upper sweep limit
$E_{\max}/h$. The low-frequency operating range and the transmon-band range are
highlighted for comparison.}
    \label{fig:Sweep}
\end{figure}

\SMsubsection{Single -parameter Fisher information}
For a known signal contrast and offset, the corresponding
single-parameter QFI is
\begin{equation}
    \mathcal{F}_Q
    =
    \frac{x^2}{4T^2\cosh^2(x/2)}.
\end{equation}
All quantities quoted in the main text are obtained using
Eq.~(\ref{eq:sm:fisher}), with
$x^{*}_{\rm shot}=2.3994$, the root of
$e^x=(x+2)/(x-2)$, and
$x^{*}_{\rm time}=5.0$ when the reset cost
$\tau_{\rm shot}\propto1/uE^3$ is included. The corresponding shot
penalties are $496\times$ at \SI{5}{\giga\hertz} and
$2.0\times10^7$ at \SI{10}{\giga\hertz}, relative to the optimum at
\SI{20}{\milli\kelvin}. More explicitly, retaining the thermal occupation factor,
\begin{equation}
    \Gamma_{1}\propto uE^{3}\coth\!\left(\frac{x}{2}\right),
\end{equation}
so that the Fisher information accumulated per unit time scales as
\begin{equation}
    \frac{\mathcal{F}_{Q}}{\tau_{\rm shot}}
    \propto
    x^{5}\frac{\coth(x/2)}{\cosh^{2}(x/2)}.
    \label{eq:time_scaling}
\end{equation}
Maximizing this expression gives
$x^{*}_{\rm time}=5.0$. At $T=\SI{20}{\milli\kelvin}$,
this corresponds to an optimal splitting of approximately  $E^{*}/{h}\simeq\SI{2.08}{\giga\hertz}$. Figure~\ref{fig:Sweep} shows the
Corresponding precision $\sigma_T$ as the upper sweep limit is increased. The uncertainty remains small when the sweep is confined to the low-frequency band but increases rapidly as the sweep is extended
toward the transmon band, demonstrating the loss of thermometric sensitivity and
parameter identifiability at high TLS energies.

\SMsection{ Tangent-space geometry}
\label{sec:tangent_geometry}

To evaluate parameter identifiability during narrow energy sweeps, we formulate a tangent-space geometric framework. We represent parameter sensitivities as data-space response vectors $\boldsymbol{v}_\theta = \partial \boldsymbol{\mu} / \partial \theta$ across sweep points $E_i$. Systematic distortions span the subspace $\mathcal{S}_{\mathrm{sys}} = \text{span}\{\boldsymbol{v}_a, \boldsymbol{v}_b, \boldsymbol{v}_\Delta, \boldsymbol{v}_{T_{1,q}}\}$, where $\boldsymbol{v}_{T_{1,q}} = \partial_{T_{1,q}}\boldsymbol{\mu}$ accounts for readout qubit relaxation drift.

Under the weighted inner product $\langle \boldsymbol{x}, \boldsymbol{y} \rangle_W = \boldsymbol{x}^\top W \boldsymbol{y}$ with noise variance weights $W_{ii} = [\mu_i(1 - \mu_i)]^{-1}$, the weighted principal angle $\theta(\boldsymbol{v}_T, \mathcal{S}_{\mathrm{sys}})$ between the thermal response $\boldsymbol{v}_T$ and the systematics subspace is defined by

\begin{equation}
\cos \theta = \max_{\boldsymbol{u} \in \mathcal{S}_{\mathrm{sys}}} \frac{|\langle \boldsymbol{v}_T, \boldsymbol{u} \rangle_W|}{\|\boldsymbol{v}_T\|_W \|\boldsymbol{u}\|_W}.
\end{equation}

\begin{figure}[t!]
  \centering
  \includegraphics[width=0.4\linewidth]{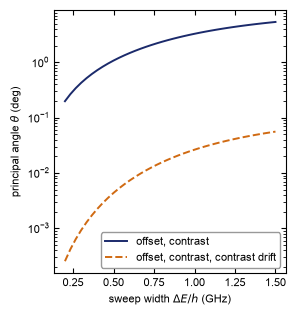}
  \caption{Weighted principal angle between the thermal response and the systematics subspace versus sweep width, for the reported defect swept upward from $\Delta_0/h=\SI{1.3}{\giga\hertz}$, for the offset--contrast subspace and with the contrast-drift direction added ($g_0/2\pi=\SI{3.0}{\mega\hertz}$, $T_{1,q}=\SI{20}{\micro\second}$).}
  \label{fig:tangent_geometry}
\end{figure}

For the reported defect, swept upward from its minimum at \SI{1.3}{\giga\hertz}, Fig.~\ref{fig:tangent_geometry} shows the angle between the thermal response and the offset--contrast subspace falling from $5.0^\circ$ at $\Delta E/h=\SI{1.39}{\giga\hertz}$ to $1.1^\circ$ at \SI{0.5}{\giga\hertz} and $0.43^\circ$ at \SI{0.3}{\giga\hertz}. Adding the contrast-drift direction $\boldsymbol{v}_{T_{1,q}}$ ($g_0/2\pi=\SI{3.0}{\mega\hertz}$, $T_{1,q}=\SI{20}{\micro\second}$) reduces the angle to $0.05^\circ$ at the full sweep width. This factor of about 100 reproduces the inflation found when $T_{1,q}$ is fitted as a fourth parameter (Sec.~\ref{sec:kappa4p}), so contrast drift must be controlled experimentally. Adding $\boldsymbol{v}_\Delta$ makes the thermal response nearly degenerate with the systematics subspace ($\theta<0.01^\circ$), so $\Delta_0$ must come from the spectroscopic hyperbola fit and not from the population fit, as in the protocol. A joint fit of sweeps taken at two readout amplitudes, with separate $(a_k,b_k)$ and a shared $T$, does not lift this degeneracy: its Fisher information is the sum of the two single-sweep Schur complements, which only doubles the shot count. Breaking the degeneracy requires a parameter shared across the sweeps, such as the offset removed by the interleaved reference of Sec.~\ref{sec:sequence}.

\SMsection{Finite-sample estimator performance and empirical coverage}
\label{sec:finite_sample}

To validate that our parameter estimation routine saturates the analytical CRB under finite shot counts, we evaluate its statistical performance via Monte Carlo simulations. For each strain-controlled energy point $E_i$ ($i = 1, \dots, M$), $N$ independent measurement shots yield $k_i \sim \text{Binomial}(N, p_i)$ excited-state counts, corresponding to observed fractions $y_i = k_i / N$. The physical excitation model is given by
\begin{equation}
p_i(a, b, T) = a + \frac{b}{1 + \exp\left(\frac{E_i}{k_B T}\right)},
\end{equation}
where $T$ is the target operating temperature, $b$ is the readout visibility, and $a$ is the un-reset residual background offset.

To prevent estimation bias caused by low shot counts or low thermal occupation, the parameter vector $\boldsymbol{\theta} = [a, b, T]^\top$ is extracted directly using a binomial maximum-likelihood estimator (MLE). The optimal parameters minimize the negative log-likelihood function:
\begin{equation}
\hat{\boldsymbol{\theta}} = \arg\min_{a, b, T} \left[ -\sum_{i=1}^{M} \left( k_i \ln p_i(a,b,T) + (N - k_i) \ln\left(1 - p_i(a,b,T)\right) \right) \right].
\end{equation}
Each fit starts from randomized values $a\in[0.005,0.05]$, $b\in[0.8,1.0]$, and $T$ within $\pm20\%$ of the true value inside the box $a \in [0, 0.1]$, $b \in [0.5, 1.0]$, and $T \in [1, 200]$~mK.
We perform $N_{\text{MC}} = 2000$ Monte Carlo trials on the headline sweep (\SIrange{1.3}{2.69}{\giga\hertz}, 11 points) at $T_{\text{true}} \in \{10, 20, 30, 50\}$~mK and $N \in \{10^4, 10^5, 10^6\}$ shots per point, together with the operating point $N^{*} = 8.42\times10^{6}$ at \SI{20}{\milli\kelvin}. Estimator performance is evaluated by the sample standard deviation
\begin{equation}
    \text{Std}(\hat{T}) = \sqrt{\langle (\hat{T} - \langle \hat{T} \rangle)^2 \rangle}
\end{equation}
and the bias $\text{Bias}(\hat{T}) = \langle \hat{T} \rangle - T_{\text{true}}$. At the operating point the spread is \SI{0.0817}{\milli\kelvin} against a CRB of \SI{0.0820}{\milli\kelvin}, the bias is $+\SI{0.001}{\milli\kelvin}$, and the coverage is $68.8\%$ at $1\sigma$ and $95.5\%$ at $2\sigma$ (Table~\ref{tab:finite_sample}). Wherever the box constraints are inactive, the spread agrees with the CRB within the Monte Carlo uncertainty of about $1.6\%$. Where they are active (last column), they act as prior information: the spread falls below the CRB and the coverage exceeds its nominal value. This occurs at low shot counts and at \SI{10}{\milli\kelvin}, where $\Delta_0>5k_BT/h$ places the defect outside its design window.
\begin{table*}[t]
\centering
\caption{Finite-sample Monte Carlo performance of the binomial MLE on the headline sweep. Theoretical marginalized CRB, empirical standard deviation $\mathrm{Std}(\hat{T})$, bias $\mathrm{Bias}(\hat{T})$, empirical coverage at $68.3\%$ ($1\sigma$) and $95.4\%$ ($2\sigma$), and the fraction of fits with an active box constraint, over $N_{\mathrm{MC}} = 2000$ trials.}
\label{tab:finite_sample}
\begin{tabular*}{\textwidth}{@{\extracolsep{\fill}}cccccccc}
\hline\hline
$T_{\mathrm{true}}$ (\si{\milli\kelvin}) & $N$ (shots/point) & CRB (\si{\milli\kelvin}) & $\mathrm{Std}(\hat{T})$ (\si{\milli\kelvin}) & $\mathrm{Bias}(\hat{T})$ (\si{\milli\kelvin}) & Cov ($1\sigma$) & Cov ($2\sigma$) & Box active \\
\hline
10.0 & $10^{4}$ & 17.9677 & 2.8977 & $-0.2829$ & 100.0\% & 100.0\% & 82\% \\
10.0 & $10^{5}$ & 5.6819 & 0.7423 & $+0.4317$ & 100.0\% & 100.0\% & 79\% \\
10.0 & $10^{6}$ & 1.7968 & 0.5520 & $+0.3987$ & 100.0\% & 100.0\% & 58\% \\
\hline
20.0 & $10^{4}$ & 2.3808 & 1.6470 & $+0.8552$ & 82.2\% & 95.5\% & 48\% \\
20.0 & $10^{5}$ & 0.7529 & 0.5698 & $+0.1864$ & 82.5\% & 97.2\% & 30\% \\
20.0 & $10^{6}$ & 0.2381 & 0.2310 & $+0.0155$ & 66.2\% & 97.3\% & 8\% \\
$\mathbf{20.0}$ & $\mathbf{8.42\times10^{6}}$ & $\mathbf{0.0820}$ & $\mathbf{0.0817}$ & $\mathbf{+0.0011}$ & $\mathbf{68.8\%}$ & $\mathbf{95.5\%}$ & $\mathbf{0\%}$ \\
\hline
30.0 & $10^{4}$ & 2.3712 & 1.8538 & $+0.6117$ & 84.2\% & 96.4\% & 36\% \\
30.0 & $10^{5}$ & 0.7498 & 0.6751 & $+0.0517$ & 71.4\% & 97.4\% & 10\% \\
30.0 & $10^{6}$ & 0.2371 & 0.2331 & $+0.0022$ & 69.3\% & 95.8\% & 0\% \\
\hline
50.0 & $10^{4}$ & 5.5510 & 4.0898 & $+0.4966$ & 78.5\% & 100.0\% & 31\% \\
50.0 & $10^{5}$ & 1.7554 & 1.7514 & $+0.1110$ & 68.8\% & 95.1\% & 0\% \\
50.0 & $10^{6}$ & 0.5551 & 0.5553 & $+0.0031$ & 68.2\% & 95.5\% & 0\% \\
\hline\hline
\end{tabular*}
\end{table*}
\SMsection{Measurement sequence and the readout offset}
\label{sec:sequence}

Equation~(\ref{eq:sm:fisher}) assumes a constant offset $a$ throughout the strain sweep. This is a property of the measurement sequence and must be specified before evaluating the bound.

The swap in Eq.~\eqref{eq:signal} of the main text requires the readout qubit to be resonant with the
defect. A naive implementation, therefore, tunes $f_q$ together with $E$ across the
sweep. The qubit's residual excitation then contributes to $a$, but this excitation
itself varies strongly with $E$: at \SI{20}{\milli\kelvin}, an equilibrium qubit has
$p_e=\SI{4.23}{\percent}$ at \SI{1.3}{\giga\hertz} and
$p_e=\SI{0.157}{\percent}$ at \SI{2.69}{\giga\hertz}. At the experimentally observed
$T_{\rm eff}\approx\SI{45}{\milli\kelvin}$ floor~\cite{PhysRevLett.114.240501},
the corresponding values are \SI{20.0}{\percent} and \SI{5.37}{\percent}.
Thus, $a$ cannot be assumed constant during the sweep.

The resulting bias has a simple origin. If the qubit and defect are at the same
temperature, the qubit excitation has the same $E$ dependence as the defect
population and is absorbed into $b$, leaving the temperature estimate unbiased.
The problem arises when their temperatures differ. For a defect at
\SI{20}{\milli\kelvin}, fitting a constant $a$ gives
\SI{23.0}{\milli\kelvin} for a qubit at \SI{25}{\milli\kelvin}
($+\SI{15}{\percent}$) and \SI{36.7}{\milli\kelvin} for a qubit at
\SI{45}{\milli\kelvin} ($+\SI{83}{\percent}$). The naive sequence therefore fails in precisely the regime of interest and biases the estimate toward the qubit temperature. Two alternative sequences remove this problem, and the protocol requires one of them.

\emph{(A) Interleaved reference.} At each strain point, alternate the swap
measurement $S_i$ with an otherwise identical measurement without the swap $R_i$. Their difference 
\begin{equation}
D_i=S_i-R_i=b\,p_{e,i},
\end{equation}
removes the unknown offset $a$ and leaves a two-parameter fit for $(b,T)$.
The additional reference shots double the binomial variance, but the reduced
number of fitted parameters can improve precision. For a residual
$a=\SI{1}{\percent}$ and active qubit reset, the statistical error is
$0.51$ times that of the three-parameter fit at equal shot count, giving a total uncertainty
of \SI{0.42}{\percent} at $\tau^*=\SI{3.9}{\hour}$. Without active reset, thermal excitation of the qubit at $T_{\rm eff}=\SI{45}{\milli\kelvin}$ increases the reference noise, raising the ratio to $1.35$ and the total uncertainty to \SI{0.67}{\percent} at $\tau^*=\SI{10.3}{\hour}$. Active reset is therefore part of the measurement sequence.

\emph{(B) Fixed-frequency readout with a parametric swap.} Keep the readout qubit at a fixed frequency and drive the swap at $|f_q-E|$. Since the qubit frequency remains fixed during the TLS sweep, $a$ stays independent of $E$, allowing the three-parameter estimator of Eq.~(\ref{eq:sm:fisher}) to be used. The main additional requirement is calibration of the parametric drive. For $f_q\approx\SI{5}{\giga\hertz}$, the drive lies at \SIrange{2.3}{3.7}{\giga\hertz}, outside the sweep band. The sideband rate is at most $0.58\,g$, set by the maximum of the Bessel factor $J_1$, resulting in a longer swap (Sec.~\ref{sec:contrast}).

We use sequence~(B) for the main error budget because it gives the more
cautious estimate. Sequence~(A) provides a direct check of the offset $a$ and should be used as a control for either implementation.

\SMsection{Power of the residual}
\label{sec:residual}

To show that the residual can distinguish a non-thermal state from a thermal
one, we quantify the statistical power of the test. As a simple
non-equilibrium model, we consider a mixture of a bath at temperature $T$
and a hot component at temperature $T_h$,
\begin{equation}
p_e=
\frac{(1-w)e^{-E/(k_BT)}
+w e^{-E/(k_BT_h)}}
{1+(1-w)e^{-E/(k_BT)}
+w e^{-E/(k_BT_h)}} ,
\label{eq:nonthermal}
\end{equation}
where $w$ is the weight of the hot component. We take
$T_h=\SI{200}{\milli\kelvin}$, generate the corresponding population data,
and fit them with the constant-$(a,b)$ thermal model. 
We then evaluate the resulting $\chi^2$ against the $5\sigma$ threshold,
\begin{equation}
\chi^2=\sum_i\frac{\left(S_i-S_i^{\rm fit}\right)^2}{\sigma_i^2},
\label{eq:sm:chisq}
\end{equation}
where $S_i$ is the measured signal, $S_i^{\rm fit}$ is the best-fit thermal
model, and $\sigma_i$ is the statistical uncertainty at strain point $i$.
For $k=8$ degrees of freedom, the $5\sigma$ detection threshold is
$\chi^2\simeq44.0$ (one-sided $\chi^2\simeq45.4$).
At $N^{*}=8.42\times10^{6}$ shots per strain point, this threshold is used here. For the precision sweep, \SIrange{1.3}{2.69}{\giga\hertz}, a hot component is nearly constant over the sweep and is largely absorbed into the offset $a$: $w=\SI{5}{\percent}$ gives $\chi^2=95$, and the minimum detectable admixture is $w_{\min}=\SI{2.7}{\percent}$. At $N=10^{6}$ the same admixture gives $\chi^2=11$, and no admixture up to \SI{40}{\percent} reaches the threshold. Extending the sweep to \SIrange{1.3}{3.22}{\giga\hertz} ($V_{\mathrm{sym}}=\SI{20}{\volt}$, or \SI{10}{\volt} along one branch) gives $\chi^2=441$ at $w=\SI{5}{\percent}$ and $w_{\min}=\SI{0.97}{\percent}$. The power therefore grows linearly with the shot count, since $\chi^2\propto N$ for a fixed distortion. Detection comes only after the distortion has biased the temperature. At $w_{\min}$ the fitted temperature is \SI{22.6}{\milli\kelvin} on the precision sweep ($+\SI{2.6}{\milli\kelvin}$, $13\%$) and \SI{21.3}{\milli\kelvin} on the extended sweep, whereas an admixture of $w=\SI{0.12}{\percent}$ ($\SI{0.085}{\percent}$ on the extended sweep) biases $T$ by $1\sigma$, \SI{0.116}{\milli\kelvin}, with $\chi^2=0.17$ ($0.46$). The residual test therefore detects gross non-thermality but cannot certify thermality at the level of the quoted precision.

The residual can also respond to instrumental effects, such as readout
contrast drift or a strain dependence of $\Delta_0$. These effects cannot be separated by the residual alone. We compare residual shapes, the components of each distortion left after projecting out the $(a,b,T)$ directions, with the weighted inner product. On the precision sweep the residual shape of a hot admixture correlates with that of a linear contrast drift at $|\cos\theta|=0.97$ and with that of a strain-dependent $\Delta_0$ at $0.78$, and contrast drift and $\Delta_0$ correlate at $0.89$. Extending the sweep to \SI{3.22}{\giga\hertz} changes these to $0.93$, $0.76$ and $0.92$, so the extension does not separate them. Therefore, a nonzero residual indicates a departure from the assumed thermal model. At the same time, the additional controls described in Sec.~\ref{sec:contrast} and the hyperbola-minimum test are needed to identify its origin.
\SMsubsection{What the residual does not test}
\label{sec:whichbath}
A zero residual shows that the defect is consistent with a Gibbs state, but it does
not identify the bath that sets its temperature. For example, a defect equilibrated
with a confined phonon mode or a photon environment above the mixing chamber can
still produce a Boltzmann population, a vanishing residual, and a well-defined
temperature different from the cryostat temperature. Thus, the overdetermination of Sec.~\ref{sec:est} provides a test of consistency with thermality rather than a determination of the bath responsible for the temperature.

For the intended application, this distinction is not necessarily a problem: the
error budget concerns the temperature of the degrees of freedom that limit qubit
coherence, which is precisely what the defect probes. The ambiguity matters when
comparing the defect temperature with an external thermometer.

The origin of the temperature can be tested by varying the bath while keeping the 
thermometer fixed. Sweeping the mixing-chamber temperature $T_{\mathrm{MXC}}$ 
across the operational range of $\SI{20}{}\text{--}\SI{100}{\milli\kelvin}$ against 
a trusted reference thermometer (such as a co-fabricated metallic resistor or 
CBT array), we require unit-slope tracking
\begin{equation}
\frac{\mathrm{d}T_{\rm TLS}}{\mathrm{d}T_{\rm MXC}}=1,
\label{eq:sm:unitslope}
\end{equation}
for a defect that remains thermally coupled to the mixing chamber. A local
nonequilibrium source would generally produce a slope different from unity,
whereas a static temperature offset would preserve the unit slope. Given the projected 
single-defect fit precision of $0.58\%$ ($\sigma_T = \SI{0.116}{\milli\kelvin}$ at 
$\SI{20}{\milli\kelvin}$), this test provides sub-millikelvin sensitivity, 
resolving static thermal offsets down to $\approx \SI{0.1}{\milli\kelvin}$ across 
the sub-100~mK regime. Comparing
multiple defects in the same and different junctions provides an additional
consistency check and can help localize a local thermal load. These are validation tests but not calibrations. A co-fabricated CBT array measures
the electron temperature of a metal film on the chip surface, not the temperature
inside the junction barrier. Therefore, disagreement between the two does not by
itself determine whether the in-barrier thermometer is biased or the barrier is
genuinely hotter. Until the unit-slope test in Eq.~(\ref{eq:sm:unitslope}) is
demonstrated, the appropriate description is a candidate primary
thermometer.

\SMsection{Defect yield}
\label{sec:yield}
A usable defect must satisfy four requirements: (i) it must be reachable
within the operating band, (ii) it must show a resolvable avoided crossing,
(iii) it must remain stable throughout the sweep, and (iv) its hyperbola
minimum must be sufficiently low. For the first three requirements, transmon defect spectroscopy reports a
junction-barrier defect density of
\SI{15}{\per\giga\hertz} per qubit and an avoided-crossing fraction of about \SI{3}{\percent} for splittings above \SI{5}{\mega\hertz}~\cite{Lisenfeld2019}.

The defect jump statistics are evaluated based on the 82-defect survey~\cite{PhysRevLett.105.177001}, where thermal cycling above \SI{20}{\kelvin} was used across 8 cooldowns to reset the TLS distribution, with approximately \SI{5}{\percent} of defects exhibiting spontaneous spectral instability. Assuming a nominal observation window of $\tau_{\mathrm{obs}} = 24\text{ h}$ per cooldown cycle, the corresponding jump rate is $\gamma_{\mathrm{jump}} \approx 2.1 \times 10^{-3}\text{ h}^{-1}$, yielding a defect survival probability of $P_{\mathrm{surv}} = e^{-\gamma_{\mathrm{jump}} \tau^*} \approx 98.4\%$ over the optimal $\tau^* = 7.63\text{ h}$ precision sweep. Across a plausible monitoring range of $\tau_{\mathrm{obs}} \in [12, 48]\text{ h}$, the survival probability remains high ($P_{\mathrm{surv}} \in [96.8\%, 99.2\%]$), confirming that defect jump instabilities contribute negligibly to the overall uncertainty budget. 

The sweep-band condition and the requirement on $\Delta_0$ are not independent instead, both follow from the same joint distribution
$P(\varepsilon,\Delta_0)$.  We therefore avoid factorizing these two
conditions and integrate directly over the accessible region of the
$(\varepsilon,\Delta_0)$ plane.  With
\begin{equation}
    E=\sqrt{\varepsilon^2+\Delta_0^2},
\qquad \text{and}\qquad
P(\Delta_0)\propto\frac{1}{\Delta_0},
\end{equation}
the normalized distribution over a population spanning
\begin{equation}
    R=\frac{\Delta_{0,\max}^{\rm pop}}{\Delta_{0,\min}^{\rm pop}}
\end{equation}
is
\begin{equation}
    p(\Delta_0)=\frac{1}{\Delta_0\ln R}.
\end{equation}
The accessible asymmetry-energy interval at fixed $\Delta_0$ for a sweep from $E_{\min}$ to $E_{\max}$ is
\begin{equation}
\begin{aligned}
    \varepsilon_{\min}(\Delta_0)
=&
\sqrt{\max(E_{\min}^2-\Delta_0^2,0)},
\\
\varepsilon_{\max}(\Delta_0)
=&
\sqrt{E_{\max}^2-\Delta_0^2}.
\end{aligned}
\end{equation}
The expected number of qualifying defects per qubit is obtained by integrating
the joint $(E,\Delta_0)$ distribution over the accessible region of the strain sweep
\begin{equation}
\lambda(R)
=
\rho_{\rm TLS} f_{\rm AC} f_{\rm st}
\int_{E_{\min}}^{E_{\max}} dE
\int_{\Delta_{0,\min}^{\rm op}}^{
\min(\Delta_{0,\max}^{\rm op},E)}
\frac{d\Delta_0}{\Delta_0\ln R},
\label{eq:sm:yield_joint}
\end{equation}
where
$R$ is the assumed population span of $\Delta_0$.  This form avoids treating the sweep-band and
$\Delta_0$ requirements as independent filters.

For the headline operating range $E/h = 1.3\text{ GHz}$ to $2.69\text{ GHz}$ and $\Delta_0/h = 1.0\text{ GHz}$ to $1.5\text{ GHz}$, the resulting yields are re-integrated over the updated energy bounds, giving $\lambda \approx 0.051$ for a two-decade population span and $\lambda \approx 0.034$ for a three-decade span. For a narrow $\Delta_0/h = 1.2\text{ GHz}$ to $1.4\text{ GHz}$ window centered on the headline defect, the corresponding yields are $\lambda \approx 0.020$ and $0.010$, respectively.  These values should be
regarded as extrapolations because both the defect density and the
avoided-crossing statistics are measured primarily in the transmon band and
are being applied here to the lower-frequency operating range.  Moreover,
the quoted \SI{3}{\percent} avoided-crossing fraction was measured for
$u\ll1$; since $g=g_0\sqrt{u}$ is maximal near the symmetry point, applying
this fraction to the present operating band is conservative and likely
underestimates the usable-defect yield. For a chip with $N_q = 10$ qubits, the Poisson relation $P_{\ge 1} = 1 - e^{-N_q \lambda}$ with $\lambda=0.034$--$0.051$ gives $P_{\ge1}=29\%$--$40\%$ for at least one usable defect and a $90\%$ chance needs 45 to 70 junctions. The avoided-crossing criterion of \SI{5}{\mega\hertz} corresponds to $g\gtrsim2\pi\times\SI{2.5}{\mega\hertz}$, consistent with the \SI{3.0}{\mega\hertz} coupling of the reported defect.

\SMsection{In-situ spectroscopic tracking and intra-dwell smearing budget}
\label{sec:sdb}
Slow spectral wander of the defect transition frequency is continuously tracked and subtracted using in-situ adaptive spectroscopy during the sweep. The remaining contribution is the tracking residual, which is the spectral variation during each dwell that tracking does not correct. It is evaluated by propagating the residual point by point through the estimator using Eq.~(\ref{eq:sm:diffprop}). We use the diffusivity measured for barrier and near-junction defects by Klimov \emph{et al.}~\cite{PhysRevLett.121.090502}. They report $D=\SI{2.5}{\mega\hertz}\,\mathrm{h}^{-1/2}$, with the rms frequency wander given by
\begin{equation}
\sigma(t)=2D\sqrt{t}.
\label{eq:sm:diff}
\end{equation}
This gives an rms wander of \SI{5.0}{\mega\hertz} after one hour, rather than \SI{2.5}{\mega\hertz}.
The factor of two in Eq.~(\ref{eq:sm:diff}) is important because omitting it
changes the inferred floor by $2.13\times$.

The same convention has since been tested independently. Reference~\cite{4ssz-6ctb}
tracks eight defects over \SI{24}{\hour} in flux-tunable transmons with different
geometries and fabrication, and obtains
$D=\SI{2.2(1)}{\mega\hertz}\,\mathrm{h}^{-1/2}$ using the same
Eq.~\eqref{eq:sm:diff} convention. This agrees with the
$\SI{2.5(1)}{\mega\hertz}\,\mathrm{h}^{-1/2}$ reported in
Ref.~\cite{PhysRevLett.121.090502} however, we retain the larger value as the more
conservative estimate. The \SI{20}{\minute} timing resolution of
Ref.~\cite{4ssz-6ctb} probes the slow component but cannot resolve the faster
dynamics discussed below.

Sub-second adaptive spectroscopy has since resolved this faster component
\cite{berritta2026adaptivespectroscopyfasttwolevelsystem}. The reported Ornstein--Uhlenbeck correlation time is $\tau_c=\SI{32}{\second}$, so the frequency wander saturates at long times. For comparison with Eq.~(\ref{eq:sm:diff}), the corresponding short-lag diffusion
constant is
\begin{equation}
    D_s=3.47\times10^{-3}\,\mathrm{MHz^2\,s^{-1}},
\end{equation}
while Ref.~\cite{berritta2026adaptivespectroscopyfasttwolevelsystem} reports
$D=\SI{0.93}{\mega\hertz^2\per\second}$ for its short-time dynamics. The two
measurements are therefore complementary: the \SI{20}{\minute} resolution of
Ref.~\cite{4ssz-6ctb} cannot resolve a \SI{32}{\second} correlation time, whereas
the $\sim\SI{e3}{\second}$ records of Ref.~\cite{berritta2026adaptivespectroscopyfasttwolevelsystem}
do not probe hour-scale drift.
\begin{table*}[t!]
\caption{\label{tab:sm:budget}
Accuracy budget at \SI{20}{\milli\kelvin}, evaluated at the sweep duration $\tau^{*}=\SI{7.63}{\hour}$ ($N_{\text{tot}} \approx 9.27 \times 10^{7}$ total sweep shots, $N^* = 8.42 \times 10^{6}$ per strain point, $\tau_{\rm shot}=\SI{0.30}{\milli\second}$ averaged over the sweep, $D = \SI{2.5}{\mega\hertz\hour^{-1/2}}$), for the reported defect with $\Delta_{0}/h=\SI{1.3}{\giga\hertz}$ swept to \SI{2.69}{\giga\hertz}, which requires an asymmetry excursion of \SI{2.36}{\giga\hertz}, or \SI{8}{\volt} along one branch. The smearing term is the tracking residual at $\eta=1$, propagated point by point with Eq.~(\ref{eq:sm:diffprop}).}
\begin{ruledtabular}
\footnotesize
\setlength{\tabcolsep}{3pt}
\begin{tabular}{@{}lcc@{}}
Contribution & $\sigma_T$ (mK) & $\sigma_T/T$ \\
\colrule
\multicolumn{3}{l}{\emph{Random uncertainty budget (quadrature sum)}}\\
Intra-dwell spectral smearing, propagated (\SI{7.63}{\hour}) & 0.0824 & \SI{0.412}{\percent}\\
Counting statistics ($8.42 \times 10^{6}$ per point) & 0.0820 & \SI{0.410}{\percent}\\
Hyperbola fit ($\Delta_0$) & 0.0060 & \SI{0.030}{\percent}\\
$E$ spectroscopy & 0.0002 & \SI{0.001}{\percent}\\
Fast bounded wander, refitted ($\sigma_f=\SI{5.5}{\mega\hertz}$) & $\le 0.0007$ & $\le\SI{0.004}{\percent}$\\
\colrule
Total random uncertainty ($\sigma_{\text{rand}}$, in quadrature) & 0.1164 mK & \SI{0.582}{\percent}\\

\colrule
\multicolumn{3}{l}{\emph{Systematic bias budget (one-sided / linear bound)}}\\
Mean-force correction & $\le+0.0012$ & $\le+\SI{0.006}{\percent}$\\
\colrule
Total primary systematic bias ($\Delta T_{\text{sys}}$) & $[0, +0.0012]$ & $[0, +\SI{0.006}{\percent}]$\\

\colrule
\multicolumn{3}{l}{\emph{Additional readout allowance: $\lvert\kappa\rvert\le0.011$}}\\
Contrast drift ($T_{1,q}\gtrsim\SI{12}{\micro\second}$ at $g_0/2\pi=\SI{3}{\mega\hertz}$) &
$\le\SI{0.043}{\milli\kelvin}$ & $\le\SI{0.215}{\percent}$\\

\colrule
Combined systematic bias bound ($\sum |\text{bias}|$) & $\le 0.044$ & $\le\SI{0.22}{\percent}$\\

\colrule
\multicolumn{3}{l}{\emph{Total Expanded Uncertainty Budget}}\\
Total uncertainty ($0.1164$ mK random $+$ biases) &
$0.1164 \text{ (rand)};\ \le 0.044 \text{ (sys)}$ &
$\SI{0.582}{\percent}\text{ (rand)};\ \le\SI{0.22}{\percent}\text{ (sys)}$\\
\end{tabular}
\end{ruledtabular}
\end{table*}

Reference~\cite{berritta2026adaptivespectroscopyfasttwolevelsystem} defines $D$ through the short-lag mean-squared displacement $2Dt$, so the stationary rms wander of the fast Ornstein--Uhlenbeck (OU) component is
\begin{equation}
    \sigma_f = \sqrt{D\tau_c} = \SI{5.5}{\mega\hertz},
    \label{eq:fast_jitter_rms}
\end{equation}
with a robustness range of \SIrange{4.0}{5.9}{\mega\hertz}. This is comparable
to the \SI{4.37}{\mega\hertz} slow-diffusion wander accumulated over one dwell, $\tau^*/10 = \SI{0.763}{\hour}$.
Both the correlation time $\tau_c = \SI{32}{\second}$ and the telegraphic switching time
$\tau_{\rm sw} = \SI{2.2}{\second}$ are much shorter than the
\SI{45.8}{\minute} spent at each strain point, so the fast fluctuations average within each point. To second order in the frequency wander,
\begin{equation}
\langle p_e(x+\delta)\rangle
=
p_e(x)+\frac{1}{2}\sigma_x^2 p_e''(x),
\end{equation}
where
\begin{equation}
\sigma_x=
\frac{\varepsilon}{E}
\frac{\sigma_f}{k_BT/h},
\qquad
\delta T=
\frac{1}{2}\sigma_x^2 p_e''(x^*)
\left(
\frac{\partial p_e}{\partial T}
\right)^{-1}_{x^*}.
\label{eq:sm:fastwander}
\end{equation}

However, this
estimate is not sufficient because $\sigma_x$ and $p_e''(x)$ vary across the
sweep. We therefore evaluate the second-order correction at each of the eleven
strain points and refit the resulting distorted sweep with the same
three-parameter estimator. This gives
\begin{equation}
\delta T=\SI{4.94e-4}{\milli\kelvin},
\label{eq:sm:fastwander-refit}
\end{equation}
rising to \SI{6.02e-4}{\milli\kelvin} at the upper
\SI{4.2}{\mega\hertz} value of the robustness range. The refitted value is used
in Table~\ref{tab:sm:budget}. It is still $135\times$ below the residual smearing floor and changes the total uncertainty only in the fourth decimal place. Even an ungated telegraphic excursion of $\pm\SI{50}{\mega\hertz}$ would contribute only \SI{0.038}{\milli\kelvin}, changing the total from \SI{0.581}{\percent} to \SI{0.612}{\percent}. Thus, the residual intra-dwell smearing floor remains the dominant systematic contribution under in-situ tracking.

There is, however, an important observation-window limitation. The records of
Ref.~\cite{berritta2026adaptivespectroscopyfasttwolevelsystem} span only
$\sim\SI{e3}{\second}$, so a component with a correlation time between that
timescale and one hour would remain unresolved and be included in the slow
diffusion estimate of Eq.~(\ref{eq:sm:diff}). The floor is therefore unchanged,
but its uncertainty is not reduced. Moreover, the values of $D$, $\tau_c$, and
$\tau_{\rm sw}$ are measured for defects in the transmon band. Applying these
timescales to the sub-\SI{2}{\giga\hertz} operating band is therefore an
extrapolation, just as for $D$ itself.

The two extrapolations enter the budget differently. The diffusivity $D$ enters
linearly and sets a substantial part of the systematic floor, so measuring it
directly in the sub-\SI{2}{\giga\hertz} band is the first experimental priority. The timescales enter only through
$\tau_c,\tau_{\rm sw}\ll\tau^*/10$, which currently holds by factors of $86$ and $1.2\times10^3$, respectively. A defect with $\tau_c$ longer than $46~\mathrm{min}$ would instead appear as a slow-shift distortion and could be
identified through the hyperbola residual discussed in End Matter. The underlying strain wander projects onto transition-energy fluctuations through the local slope $dE/d\varepsilon = \varepsilon/E$ of the hyperbolic dispersion. This slope varies across the sweep, from zero at the hyperbola minimum to $0.875$ at $2.69~\mathrm{GHz}$, so we propagate it point by point through Eq.~\eqref{eq:sm:diffprop} rather than through a single operating-point factor. The systematic diffusion floor thus incorporates a geometric factor $\varepsilon/E = \sqrt{3}/2 \approx 0.866$, which projects strain-induced noise measured on asymptotic reference defects ($\varepsilon/E \approx 1$ at $\SI{5.5}{\giga\hertz}$) onto the local dispersion slope of the budget defect at $\SI{1.0}{\giga\hertz}$:
\begin{equation}
\frac{\mathrm{d}E}{\mathrm{d}\varepsilon} = \frac{\varepsilon}{\sqrt{\varepsilon^2 + \Delta_0^2}} = \frac{\varepsilon}{E} = \sqrt{1 - \left(\frac{\Delta_0}{E}\right)^2} = \frac{\sqrt{3}}{2} \approx 0.866 .
\label{eq:sm:eps_over_E}
\end{equation}
The resulting energy wander is $\sigma_E(\tau) = (\varepsilon/E) \, h \, \sigma(\tau)$. Propagating this fluctuation into temperature error yields the systematic floor:
\begin{equation}
\sigma_T^{\mathrm{diff}}(\tau)
=
\left(\frac{\varepsilon}{E}\right)
\frac{\sigma(\tau)}{E/h} \, T_0 .
\label{eq:sm:sig_diff_final}
\end{equation}
This local hyperbola slope reduces the systematic temperature floor by a factor of $E/\varepsilon = 2/\sqrt{3} \approx 1.155$ ($1/0.866$) relative to unprojected frequency diffusion. The $\tau^{1/2}$ growth of this systematic floor arises directly from the time-budget allocation: for a total sweep duration $\tau$ partitioned across $11$ strain points ($10$ step intervals), the dwell duration at each point scales as $\tau/10$. Under slow spectral diffusion, the frequency variance accumulated within a single dwell window grows as $\sigma^2(\tau/10) = 4D^2 (\tau/10) \propto \tau$. In the absence of frequency tracking, slow spectral diffusion enters at first order in temperature bias via Eq.~(\ref{eq:sm:sig_diff_final}). However, because the in-situ tracking protocol subtracts the mean frequency estimated during each dwell window, the residual intra-dwell offset has zero mean ($\langle \delta \rangle = 0$). The leading linear term in the expansion therefore cancels out, forcing the residual slow diffusion to enter the systematic temperature floor at second order $\mathcal{O}(\sigma_x^2)$ through the exact same curvature expression derived in Eq.~(\ref{eq:sm:fastwander}). While in-situ tracking subtracts inter-dwell drift between strain points, this intra-dwell diffusion remains unmitigated. Thus, counting statistics improve as $\tau^{-1/2}$, whereas the intra-dwell diffusion contribution grows as $\tau^{1/2}$. The total uncertainty can therefore be written as
\begin{equation}
\sigma_T^2(\tau)=\frac{A}{\tau}+B\tau,
\label{eq:sm:budget}
\end{equation}
which has a shallow optimum at
\begin{equation}
\tau^*
=
\tau_{\rm ref}
\frac{\sigma_T^{\rm stat}(\tau_{\rm ref})}
{\sigma_T^{\rm diff}(\tau_{\rm ref})},
\qquad
\sigma_T^{\min}
=
\sqrt{
2\,\sigma_T^{\rm stat}(\tau_{\rm ref})
\sigma_T^{\rm diff}(\tau_{\rm ref})
}.
\label{eq:sm:optimum}
\end{equation}

The resulting precision budget is summarized in
Table~\ref{tab:sm:budget}. The five terms above are intrinsic to the protocol and are what the \SI{0.58}{\percent} of the main text quotes. The readout-contrast drift is a requirement on the apparatus, so it
is listed separately: it enters only to the extent that the readout fails to meet the tolerance derived in
Sec.~\ref{sec:contrast}. The $E_i$ values are treated as spectroscopically resolvable quantities, while
$\Delta_0$ is obtained from the hyperbola fit; their uncertainties are included
separately in Table~\ref{tab:sm:budget}. The hyperbola residuals independently
test the assumed strain invariance of $\Delta_0$.

\emph{Point-by-point propagation of the intra-dwell wander.} Equation~(\ref{eq:sm:sig_diff_final}) evaluates the residual wander at a single operating point. That form is adequate when the sweep is wide and the operating point sits away from the hyperbola minimum, but it fails in two regimes. First, for a defect swept from its own minimum, the local slope $\varepsilon/E$ vanishes there, so the single-point expression returns a zero floor and the optimum of Eq.~(\ref{eq:sm:optimum}) is undefined. Second, it omits the amplification of energy errors by the same ill-conditioning that governs the statistical bound, so it underestimates the floor for narrow sweeps. We therefore propagate the wander point by point. Each dwell of length $\tau/(N_{\rm pts}-1)$ leaves a residual rms frequency error $\sigma_{\nu}$, which enters point $i$ through the local slope
\begin{equation}
\sigma_{E,i}=\Bigl(\frac{\varepsilon_i}{E_i}\Bigr)\,h\,\sigma_{\nu}\!\left(\frac{\tau}{N_{\rm pts}-1}\right),
\label{eq:sm:sigmaEi}
\end{equation}
and shifts the predicted signal by $\delta\mu_i=(\partial\mu_i/\partial E_i)\,\delta E_i$ with $\partial\mu_i/\partial E_i=-b\,p_{e,i}(1-p_{e,i})/k_{B}T$. The weighted least-squares estimator maps such a perturbation onto the parameters through its influence function $(J^{\mathsf T}WJ)^{-1}J^{\mathsf T}W$, with the Jacobian $J$ and the binomial weights $W$ of Eq.~(\ref{eq:sm:fisher}), so that
\begin{equation}
\bigl[\sigma_T^{\rm diff}\bigr]^{2}=\sum_{i}\Bigl\{\bigl[(J^{\mathsf T}WJ)^{-1}J^{\mathsf T}W\bigr]_{T,i}\,\frac{\partial\mu_i}{\partial E_i}\,\sigma_{E,i}\Bigr\}^{2}.
\label{eq:sm:diffprop}
\end{equation}
Equation~(\ref{eq:sm:diffprop}) reduces to the single-point expression when the slope and the per-point sensitivity are flat across the sweep, and it is finite at the hyperbola minimum because the points away from the minimum carry the information. For a \SI{0.5}{\giga\hertz} defect swept to \SI{1.6}{\giga\hertz} it raises the total from 0.57\% to 0.74\% and shortens the optimum from \SI{9.18}{\hour} to \SI{5.5}{\hour}. All numbers quoted in this Supplemental Material use Eq.~(\ref{eq:sm:diffprop}).

The size of the residual is fixed by the tracking protocol, which an experiment must specify and report. A residual $\eta\,\sigma_\nu$ gives $\sigma_T^{\min}\propto\eta^{1/2}$ and $\tau^{*}\propto\eta^{-1}$: $\eta=1/\sqrt{3}$ (one reference per dwell) gives $0.44\%$ at $\tau^{*}=\SI{13.1}{\hour}$, and $\eta=1/\sqrt{12}$ (references at both ends) gives $0.31\%$ at \SI{26.2}{\hour}. Exact removal of the dwell mean, $\eta\to0$, would leave only the second-order smearing of Eq.~(\ref{eq:sm:fastwander}), and the budget would become statistics-limited with no optimum $\tau^{*}$. Tracking also costs measurement time: spending a fraction $f$ of the run on it scales the precision by $(1-f)^{-1/4}$, so $f=10\%$ gives $0.60\%$. Finally, the equal-shot allocation makes the dwell time differ by a factor of $1.9$ across the sweep; using the actual per-point dwell in place of $\tau/10$ changes the total from $0.581\%$ to $0.579\%$.

\emph{Scope and limitations of the uncertainty budget in Table~\ref{tab:sm:budget}.} Every entry is a
single-realisation error for \emph{one} defect swept once: it is a fit precision, of the
same kind as the \SI{0.35}{\percent} quoted for CBT in Table~\ref{tab:cmp} of the End Matter, and not
a metrological uncertainty. Three contributions that a metrological budget would carry
are absent, because none of them can be estimated from the present literature:
reproducibility across defects in one junction, across devices, and across cooldowns.
For TLSs the first of these is not obviously small, because $\Delta_{0}$, $\gamma_{S}$, $g_{0}$
and the local phonon environment all vary by orders of magnitude across the population. A
protocol that returns \SI{1}{\percent} on one defect and disagrees with the next defect
in the same barrier by \SI{10}{\percent} is not a \SI{1}{\percent} thermometer. Fixing the size of that term requires a measurement, which is why the main text identifies inter-comparison, rather than fit precision as the validation target. The quoted precision is what the protocol can achieve on a single defect. This determines whether the experiment is worth attempting and sets an upper bound on the accuracy ultimately achievable.

The diffusivity in Eq.~(\ref{eq:sm:diff}) was measured for defects near \SI{5.5}{\giga\hertz}. Its extrapolation to the sub-\SI{2}{\giga\hertz} range required by the protocol remains untested. Measuring it should be the first experimental step. The budget becomes statistics-limited once the rms wander falls
below \SI{2.05}{\mega\hertz\per\sqrt\hour}, a factor \num{2.4} below the
measured value. Because $\sigma_{T}^{2}=A/\tau+B\tau$ with
$B\propto D^{2}$, the optimum obeys
\begin{equation}
\sigma_{T}^{\min}=\sqrt{2}\,(AB)^{1/4}\propto D^{1/2},
\qquad
\tau^{*}=\sqrt{A/B}\propto D^{-1}.
\label{eq:sm:dscaling}
\end{equation}
Thus, the precision depends only weakly on the assumed phonon-diffusion extrapolation, with $D$ entering the budget. The same optimization gives $\sigma_T^{\min}\propto T_1^{1/4}$ and
$\tau^*\propto T_1^{1/2}$, because $\tau_{\mathrm{shot}}=m/\Gamma_1\propto T_1$.
Therefore, increasing $T_1$ by two orders of magnitude worsens the precision by a factor of $100^{1/4}\simeq3.2$, while increasing the optimum averaging time by a factor of $100^{1/2}=10$. A factor-$100$ increase in $T_1$ therefore gives a precision of approximately \SI{1.8}{\percent} and pushes $\tau^*$ to roughly \SI{76}{\hour}. Using the independently measured $D = \SI{2.2}{\mega\hertz\hour^{-1/2}}$ gives a precision of $0.54\%$ at $\tau^* = 8.7~\mathrm{h}$. If the diffusivity in the operating band is twice the Klimov value, the precision becomes $0.82\%$ at $3.8~\mathrm{h}$. At four times that value, it becomes $1.16\%$ at $1.9~\mathrm{h}$. A suppressed phonon density of states at the operating frequency would therefore degrade protocol throughput via the $T_1^{1/4}$ scaling derived above, while simultaneously weakening the low-frequency continuum extrapolation. Only a diffusivity an order of magnitude above the transmon-band value would put the protocol outside the range where it competes with the entries of Table~I.

\begin{figure}[b]
\centering
\includegraphics[width=\linewidth]{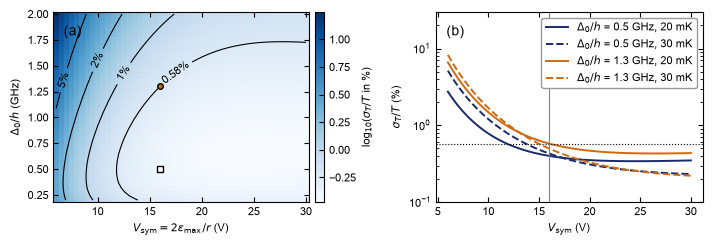}
\caption{(a) Attainable $\sigma_T/T$ at $20\,\mathrm{mK}$ over tunneling energy and symmetric strain range, propagated model, with contours at $0.58\%$, $1\%$, $2\%$, and $5\%$. The circle marks the recommended configuration ($\Delta_0 = 1.3\,\mathrm{GHz}, V_{\mathrm{sym}} = 16\,\mathrm{V}$) and the square a \SI{0.5}{\giga\hertz} comparison defect at the same voltage. (b) Precision against strain range for both defects at $20$ and $30\,\mathrm{mK}$; dotted curves show the single-point formula at $20\,\mathrm{mK}$ for comparison. Both curves flatten above about $20\,\mathrm{V}$, where the systematic floor takes over from counting statistics.}
\label{fig:phase_diagram}
\end{figure}

\SMsection{ Voltage \& defect sensitivity analysis}
To justify the choice of operational parameters ($V_{\mathrm{sym}} = 16\,\mathrm{V}$ and $\Delta_0 = 1.3\,\mathrm{GHz}$), we perform a systematic sensitivity analysis across the symmetric strain voltage $V_{\mathrm{sym}}$ and tunneling splitting $\Delta_0$. Table~\ref{tab:error_budget} summarizes the resulting relative precision $\sigma_T/T$ evaluated at $T = 20\,\mathrm{mK}$, $30\,\mathrm{mK}$, and $50\,\mathrm{mK}$. Here $V_{\rm sym}=2\varepsilon_{\max}/r$ is the symmetric-equivalent range; a one-branch sweep needs $V_{\rm sym}/2$, listed in the fourth column. Each entry quotes the point-by-point propagated budget. The single-operating-point formula is not shown, because it returns an artificially small floor whenever the sweep starts at the hyperbola minimum, where $\partial E/\partial\varepsilon\to0$. The two-dimensional parameter landscape at $T = 20\,\mathrm{mK}$ is illustrated in Figure~\ref{fig:phase_diagram}(a). Contour levels demonstrate a broad performance valley where relative errors drop below $1\%$. As shown in Figure~\ref{fig:phase_diagram}(b), increasing $V_{\mathrm{sym}}$ initially yields rapid improvements in precision by widening the available spectral sweep window. Above $V_{\mathrm{sym}} \approx 20\,\mathrm{V}$ the precision curves flatten, because the tracking residual of the TLS spectral diffusion takes over from counting statistics. The selected configuration, $V_{\mathrm{sym}} = 16\,\mathrm{V}$ (\SI{8}{\volt} along one branch) and $\Delta_0 = 1.3\,\mathrm{GHz}$, lies near the knee of this curve at $\sigma_T/T = 0.58\%$. A \SI{0.5}{\giga\hertz} defect would do better at the same voltage ($0.40\%$); the \SI{1.3}{\giga\hertz} defect is chosen because it is the one reported.
\begin{table*}[t!]
\centering
\caption{Sensitivity of the propagated budget to strain range and defect. Relative precision $\sigma_T/T$ at three temperatures; each entry re-optimizes the total sweep duration $\tau^{*}$.}
\label{tab:error_budget}
\vspace{0.2cm}
\begin{tabular*}{\textwidth}{@{\extracolsep{\fill}}ccccccc}
\hline\hline
$V_{\mathrm{sym}}$ (V) & Defect & Sweep range (GHz) & One branch (V) & $20\,\mathrm{mK}$ & $30\,\mathrm{mK}$ & $50\,\mathrm{mK}$ \\
\hline
$6.3$ & $\Delta_0 = 0.5\,\mathrm{GHz}$ & $0.5 - 1.06$ & $3.15$ & $2.40\%$ & $4.44\%$ & $11.75\%$ \\
$6.3$ & $\Delta_0 = 1.3\,\mathrm{GHz}$ & $1.3 - 1.60$ & $3.15$ & $5.68\%$ & $7.07\%$ & $13.59\%$ \\
$10.3$ & $\Delta_0 = 0.5\,\mathrm{GHz}$ & $0.5 - 1.60$ & $5.15$ & $0.74\%$ & $1.12\%$ & $2.59\%$ \\
$10.3$ & $\Delta_0 = 1.3\,\mathrm{GHz}$ & $1.3 - 2.00$ & $5.15$ & $1.37\%$ & $1.54\%$ & $2.74\%$ \\
$16.0$ & $\Delta_0 = 0.5\,\mathrm{GHz}$ & $0.5 - 2.41$ & $8.00$ & $0.40\%$ & $0.43\%$ & $0.77\%$ \\
$\mathbf{16.0}$ & $\mathbf{\Delta_0 = 1.3\,\mathrm{GHz}}$ & $\mathbf{1.3 - 2.69}$ & $\mathbf{8.00}$ & $\mathbf{0.58\%}$ & $\mathbf{0.50\%}$ & $\mathbf{0.74\%}$ \\
$20.0$ & $\Delta_0 = 0.5\,\mathrm{GHz}$ & $0.5 - 2.99$ & $10.00$ & $0.35\%$ & $0.31\%$ & $0.46\%$ \\
$20.0$ & $\Delta_0 = 1.3\,\mathrm{GHz}$ & $1.3 - 3.22$ & $10.00$ & $0.47\%$ & $0.33\%$ & $0.41\%$ \\
\hline\hline
\end{tabular*}
\end{table*}

\SMsection{Shot budget and duty cycle}
Throughout the analysis, we use $N^* \approx 8.42 \times 10^6$ shots per point at $\tau^* = \SI{7.63}{\hour}$ and $11$ strain points ($N_{\text{tot}}^* \approx 9.27 \times 10^7$ total shots). This corresponds to a shot duration
\begin{equation}
\tau_{\rm shot}
=\frac{\tau^*}{N_{\text{tot}}^*}
=\frac{\tau^*}{N_{\rm pts}N^*}
=\SI{3.27}{\milli\second},
\end{equation}
which is independent of $\tau$ because $N^*\propto\tau^*$. This duration is
consistent with the estimated defect lifetime in the required frequency
range, although it is longer than the microsecond lifetimes measured at
higher splittings.

At the symmetry point, reported lifetimes are
$\SIrange{0.44}{3.2}{\micro\second}$ for defects with
$\Delta_{0}/h=\SIrange{6.2}{7.34}{\giga\hertz}$
\cite{Lisenfeld2016,PhysRevB.92.035442}. Rescaling the same
$\Gamma_{1}\propto uE^{3}\coth(E/2k_{B}T)$ dependence used for the
per-unit-time optimum gives
$\tau_{\rm shot}=5/\Gamma_{1}$ between
$\SI{0.01}{\milli\second}$ and $\SI{10.0}{\milli\second}$ over
$\SIrange{1.3}{2.69}{\giga\hertz}$. Thus the sweep-averaged 0.30 ms lies within the expected range. The reset time
therefore contributes to the total sweep time and cannot be treated as a
negligible overhead.
\SMsubsection{Reset completeness is an $E$-dependent contrast}
\label{sec:reset}
A further constraint is that the defect must be sufficiently reset between
shots. The reset probability is
\begin{equation}
\eta(E)=1-e^{-\tau_{\rm shot}\Gamma_{1}(E)},
\label{eq:sm:eta}
\end{equation}
where $\Gamma_{1}\propto uE^{3}\coth(E/2k_{B}T)$ varies across the sweep.
Therefore, a fixed $\tau_{\rm shot}$ produces an $E$-dependent reset
probability, which acts as an additional contrast variation and cannot in
general be absorbed into the constant parameter $b$.

With a fixed $\tau_{\rm shot}=\SI{0.30}{\milli\second}$, the shortest reported symmetry-point lifetime,
\SI{0.44}{\micro\second} at \SI{6.2}{\giga\hertz}, gives
$T_{1}=\SIrange{44}{23}{\micro\second}$ across the budget sweep.
The resulting reset probability is $\eta=0.9989\to1.0000$, giving a negligible
bias of $+\SI{0.10}{\percent}$. The longest reported lifetime,
\SI{3.2}{\micro\second} at \SI{7.34}{\giga\hertz}, gives
$T_{1}=\SIrange{0.53}{0.28}{\milli\second}$ and
$\eta=0.43\to0.66$, producing a much larger bias of
$+\SI{18.6}{\percent}$. Thus the duty cycle cannot simply be fixed in
advance.
We set the shot duration from the local relaxation rate,
\begin{equation}\label{eq:reset}
\tau_{{\rm shot},i}=\frac{m}{\Gamma_{1}(E_i)}.
\end{equation}
This makes
\begin{equation}
\eta_i=1-e^{-m}
\end{equation}
constant across the sweep, so the reset factor is absorbed into the contrast
parameter $b$.
The baseline analytical protocol considered in the main text adopts $m = 5$, corresponding to a mean $\tau_{\text{shot}} = 0.30\text{ ms}$ across the sweep and an optimal sweep duration of $\tau^* = 7.63\text{ h}$, for which the attainable statistical temperature uncertainty is $\sigma_T / T = 0.58\%$, corresponding to $\sigma_T = 0.116\text{ mK}$ at $20\text{ mK}$.
As a robustness check, we also consider the more restrictive choice
$m=10$, which doubles the mean per-shot reset time to \SI{0.59}{\milli\second}. Re-optimizing the total acquisition time gives $0.69\%$ at $\tau^{*}=\SI{10.8}{\hour}$ for the lifetime calibration of the budget. Bracketing the defect lifetime with the two reported values gives
\begin{equation}
\begin{aligned}
\text{short lifetime:}&\quad
0.50\%\ (\tau^{*}=\SI{5.5}{\hour}),\ m=5;\quad 0.59\%\ (\SI{7.9}{\hour}),\ m=10,\\[2pt]
\text{long lifetime:}&\quad
0.93\%\ (\SI{19.4}{\hour}),\ m=5;\quad 1.10\%\ (\SI{27.4}{\hour}),\ m=10.
\end{aligned}
\label{eq:sm:resetbracket}
\end{equation}
Doubling the reset overhead therefore costs about $0.1$ to $0.2$ percentage points.

These values depend on the experimentally observed defect. In particular, $\Gamma_1$ is extrapolated from $u = 1$ and $E/h = \SI{6.2}{\giga\hertz}$ to the operating band (\SIrange{1.3}{2.69}{\giga\hertz}), where $uE^3$ is smaller by factors of 52 to 108. The relevant $T_{1}$ should therefore be measured at the hyperbola minimum before the sweep, not inferred from the high-frequency data.

The choice $m=10$ also makes the protocol less sensitive to changes in
$T_{1}$ during a multi-hour sweep. The reset prescription
in Eq.~\eqref{eq:reset} therefore requires an
$E$-resolved $T_{1}$ calibration before the thermometry sweep; otherwise,
strain-dependent $T_{1}$ drift produces an $E$-dependent reset contrast
that cannot be absorbed into the single contrast parameter $b$. At $m=5$, doubling $T_{1}$ produces a $-\SI{1.4}{\percent}$ bias, and tripling it gives
$-\SI{3.6}{\percent}$. These biases fall to
$-\SI{0.03}{\percent}$ and below $\SI{0.25}{\percent}$ for $m=8$ and
$m=10$, respectively. Increasing $m$ from $5$ to $10$ raises the statistical uncertainty only as $m^{1/4}$, from $\SI{0.58}{\percent}$ to $\SI{0.69}{\percent}$ for the fast branch and from $\SI{1.48}{\percent}$ to $\SI{1.76}{\percent}$ for the slow branch.

\SMsection{Strain range and voltage calibration} 
Reaching $E_{\rm max}$ from a tunneling minimum at $\Delta_0$ requires an asymmetry excursion $\varepsilon = \sqrt{E_{\rm max}^2 - \Delta_0^2}$. Along one branch of the hyperbola this needs a voltage excursion $\varepsilon/r$, with $r = \SI{295}{\mega\hertz\per\volt}$ the measured tuning rate of the asymmetry of the $\Delta_0/h = \SI{1.3}{\giga\hertz}$ defect~\cite{Lisenfeld2015}; a symmetric sweep over both branches needs
\begin{equation}
V_{\rm sym} = \frac{2\varepsilon}{r}.
\end{equation}
For $E_{\rm max}/h=\SI{2.69}{\giga\hertz}$, $\varepsilon/h = \SI{2.36}{\giga\hertz}$, so one branch needs \SI{8.0}{\volt} and both need \SI{16}{\volt}. The symmetry point of this defect lies at \SI{-14.05}{\volt}~\cite{Lisenfeld2015}. The one-branch sweep therefore runs from \SI{-14.05}{\volt} to about \SI{-6}{\volt}, inside the range of the reported measurements, whereas a symmetric sweep would need \SI{-22}{\volt} to \SI{-6}{\volt}, beyond them. We use the one-branch sweep; sweeping both branches doubles the voltage range and adds the evenness test of the symmetry properties in the main text. The extended sweep to \SI{3.22}{\giga\hertz} needs $\varepsilon/h=\SI{2.95}{\giga\hertz}$, that is \SI{10}{\volt} along one branch.
\begin{table}[b]
\caption{\label{tab:sm:drift}
Bias in the recovered temperature caused by a linear readout-contrast drift across the \SIrange{1.3}{2.69}{\giga\hertz} sweep.}
\begin{ruledtabular}
\begin{tabular}{cc}
$\kappa$ & $\delta T/T$\\
\colrule
0.003 & \SI{0.059}{\percent}\\
0.011 & \SI{0.215}{\percent}\\
0.030 & \SI{0.593}{\percent}\\
\end{tabular}
\end{ruledtabular}
\end{table}

\SMsection{Readout-contrast drift along the sweep}
\label{sec:contrast}

Equation~(\ref{eq:sm:fisher}) assumes that the offset $a$ and contrast $b$ are constant across the strain sweep. This assumption must therefore be checked. The transverse weight $u = (\Delta_0/E)^2$ decreases from $1$ at the hyperbola minimum ($\Delta_0/h = \SI{1.3}{\giga\hertz}$) to $0.233$ at $E/h = \SI{2.69}{\giga\hertz}$. Since $g = g_0\sqrt{u}$, the swap time increases by a factor of $2.07$ across the sweep. Relaxation of the readout qubit during the swap can therefore make the readout contrast, and hence the fitted parameter $b$, depend on $E$. A barrier defect tuned through a readout resonator can also strongly reduce the readout signal~\cite{Lisenfeld2016}.

We model a small contrast drift as
\begin{equation}\label{eq:sm:drift}
b(E) = b_0 \left[1 + \frac{\kappa (E - \bar{E})}{\bar{E}}\right],
\end{equation}
generate the signal with this $E$-dependent contrast, and fit it with the constant-$b$ three-parameter model. The resulting temperature bias is approximately linear in $\kappa$. We adopt $|\kappa| \le 0.011$ as the contrast-drift allowance, corresponding to a temperature bias of about $0.22\%$ over the full sweep. This allowance is kept separate from the intrinsic $0.58\%$ statistical precision budget and is included only as an apparatus requirement. The numerical results are shown in Table~\ref{tab:sm:drift}.

The contrast requirement can be related directly to the readout-qubit lifetime. The swap time is
\begin{equation}
t_{\rm swap} = \frac{\pi}{2g}, \quad g = g_0\sqrt{u}.
\end{equation}
For the coupling $g_0/2\pi = \SI{3.0}{\mega\hertz}$ reported for this defect~\cite{Lisenfeld2015}, the swap time increases from \SI{83}{\nano\second} at the hyperbola minimum to \SI{172}{\nano\second} at \SI{2.69}{\giga\hertz} (from \SI{4.2}{\nano\second} to \SI{8.6}{\nano\second} at the \SI{60}{\mega\hertz} assumed in earlier versions). The visibility contains the factor
\begin{equation}
b(E) = b_0 \exp\left[-\frac{t_{\rm swap}(E)}{T_{1,q}}\right].
\end{equation}
Requiring the adopted allowance $|\kappa| \le 0.011$ corresponds to approximately
\begin{equation}
T_{1,q} \ge \SI{11.6}{\micro\second}.
\end{equation}
A parametric swap runs at most at about $0.58\,g$, the maximum of the Bessel factor $J_1$, which raises the requirement to about \SI{20}{\micro\second}. Both values are within reach of modern transmons and fluxonium. The phase qubits of the earlier barrier-defect measurements, with $T_{1,q} \approx \SI{100}{\nano\second}$~\cite{Lisenfeld2016}, could not read this defect out by swap, because the swap alone lasts \SIrange{83}{172}{\nano\second}.

\SMsubsection{The drift must be controlled, not fitted}
\label{sec:kappa4p}
The linear model in Eq.~(\ref{eq:sm:drift}) is only a convenient way to quantify
the tolerance. The physical contrast is nonlinear because
$t_{\rm swap}\propto1/\sqrt{u}$. Using the full visibility
\begin{equation}
b(E)=b_{0}\exp\!\left[-\frac{t_{\rm swap}(E)}{T_{1,q}}\right]
\end{equation}
gives, at $g_0/2\pi=\SI{3.0}{\mega\hertz}$, temperature biases of $-20\%$ at $T_{1,q} = \SI{100}{\nano\second}$, $-2.4\%$ at \SI{1}{\micro\second}, $-0.25\%$ at \SI{10}{\micro\second} and $-0.12\%$ at \SI{20}{\micro\second} ($-1.2\%$, $-0.30\%$ and $-0.012\%$ at \SI{100}{\nano\second}, \SI{0.42}{\micro\second} and \SI{10}{\micro\second} for \SI{60}{\mega\hertz}). It requires $T_{1,q} \ge \SI{11.6}{\micro\second}$ for the $|\kappa| \le 0.011$ allowance, the same as the linear model.

Fitting the contrast drift as an additional free parameter is not an effective
alternative. Promoting $T_{1,q}$ to a fourth fit parameter makes the Fisher
matrix nearly singular because the contrast variation and thermal population
have similar dependence on $E$. At $T_{1,q}=\SI{20}{\micro\second}$ and $g_0/2\pi=\SI{3.0}{\mega\hertz}$, the statistical uncertainty increases by a factor of about 100, so the counting term alone would rise from $0.41\%$ to about $40\%$. The contrast should therefore be controlled experimentally, not fitted.

An \emph{in situ} check is to repeat the sweep at different readout amplitudes
and verify that the extracted temperature is unchanged. This directly tests
the readout assumption without calibrating the thermometer's energy axis.
\SMsection{Mean-force Gibbs correction}
\label{sec:mfgs}

Equation~\eqref{eq:signal} of the main text assumes that the defect is in the bare Gibbs state of $H_{\mathrm{TLS}}$. However, a system coupled to a bath at finite strength relaxes to the mean-force Gibbs state~\cite{PhysRevLett.127.250601},
\begin{equation}
\tau_{\mathrm{MF}}
=
\frac{\mathrm{Tr}_{B}\,e^{-\beta H}}
{\mathrm{Tr}\,e^{-\beta H}},
\label{eq:sm:mfgs}
\end{equation}
where
\begin{equation}
 H=H_{\mathrm{TLS}}+H_{B}+\tau_{z}\otimes\hat{B}.
\end{equation}
Equation~\eqref{eq:sm:mfgs} differs from $e^{-\beta H_{\mathrm{TLS}}}/Z$ at finite coupling. Because the
protocol is intended to be candidate primary, this correction is important: a
coupling-dependent distortion of $p_e(E)$ could otherwise introduce a
material-dependent parameter.
\begin{figure*}[t]
\includegraphics[width=\textwidth]{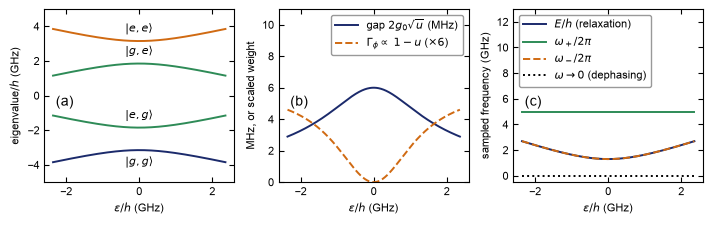}
\caption{Eigenvalue spectrum and the frequency bands each level
contributes. (a) Four-level TLS--qubit spectrum versus strain
($\Delta_{0}/h=\SI{1.3}{\giga\hertz}$, $g_{0}/2\pi=\SI{3.0}{\mega\hertz}$, $\omega_{q}/2\pi=\SI{5.0}{\giga\hertz}$). (b) Avoided-crossing gap $2g_{0}\sqrt{u}$
and pure dephasing $\propto1-u$ (the Gaussian channel, which is the measured echo form;
see text): readout is strongest where dephasing vanishes.
(c) Frequencies at which the environment is sampled: the splitting $E/h$
(relaxation), $\omega\!\to\!0$ (dephasing), and the dressed branches $\omega_{\pm}$ of Eq.~\eqref{eq:sm:hybridized_frequencies}}\label{fig:4}
\end{figure*}
\SMsubsection{Second-order result}

Expanding Eq.~(\ref{eq:sm:mfgs}) to second order in the system--bath coupling gives
\begin{equation}
\mathrm{Tr}_{B}e^{-\beta H}
=
Z_{B}e^{-\beta H_{\mathrm{TLS}}}
\left[1+K+\mathcal{O}(\lambda^{4})\right]
\end{equation},
with
\begin{equation}
K=
\int_{0}^{\beta}d\tau_{1}
\int_{0}^{\tau_{1}}d\tau_{2}\,
\tilde{A}(\tau_{1})\tilde{A}(\tau_{2})
C(\tau_{1}-\tau_{2}),
\end{equation}
where
\begin{equation}
\tilde{A}(\tau)
=
e^{\tau H_{\mathrm{TLS}}}\tau_{z}e^{-\tau H_{\mathrm{TLS}}}
\end{equation}
and
\begin{equation}
 C(\tau)
=
\langle\tilde{B}(\tau)\tilde{B}(0)\rangle_{B}
=
\int_{0}^{\infty}d\omega\,J(\omega)
\left[(n+1)e^{-\omega\tau}+ne^{\omega\tau}\right].
\end{equation}
In the energy eigenbasis, the diagonal elements are
\begin{equation}
 K_{mm}=
\sum_{k}|A_{mk}|^{2}I(E_m-E_k),
\end{equation}
where
\begin{align}
I(\Delta)
&=
\int_{0}^{\beta}ds\,(\beta-s)e^{\Delta s}C(s)
\notag\\
&=
\int_{0}^{\infty}d\omega\,J(\omega)
\left[(n+1)g(\Delta-\omega)
+n\,g(\Delta+\omega)\right],
\end{align}
with
\begin{equation}
g(z)=\frac{e^{\beta z}-1-\beta z}{z^{2}},
\qquad
g(0)=\frac{\beta^{2}}{2}.
\end{equation}

Using Eq.~(\ref{eq:sm:rot}),
\begin{equation}
|A_{ee}|^{2}=|A_{gg}|^{2}=1-u,
\qquad
|A_{eg}|^{2}=u,
\end{equation}
which separates the longitudinal and transverse contributions to the TLS dynamics, as illustrated in Fig.~\ref{fig:4}. The population correction is therefore
\begin{equation}
\begin{aligned}
\delta p_{e}
&=
p_{e}(1-p_{e})\,u\,[I(E)-I(-E)],\\[2pt]
\frac{\delta T}{T}
&=
\frac{u[I(E)-I(-E)]}{x}.
\end{aligned}
\label{eq:sm:key}
\end{equation}
Thus, only the transverse weight $u$ shifts the populations. We therefore compute
only the diagonal elements $K_{mm}$ for the temperature bias.

Reference~\cite{PhysRevLett.127.250601} also shows that the second-order
mean-force state can retain coherences in the energy eigenbasis through terms
coupling eigenoperators at different Bohr frequencies. Here, these frequencies are
$0$ and $\pm E$, so such terms are generally nonzero. They do not enter the present
budget because the swap readout of Eq.~\eqref{eq:signal} of the main text measures the population, not the energy-basis coherences. The mean-force correction is therefore controlled by the same internal coordinate
$u$ that already determines the relative weights of relaxation, dephasing, and
readout in the main text. It is largest near the degeneracy point, where $u$ is largest.

\SMsubsection{Calibration of the bath}
We assume a super-Ohmic (Debye) spectrum, $J(\omega)=A\omega^{3}$, consistent with the relaxation rate $\Gamma_{1}\propto uE^{3}\coth(E/2k_BT)$. We determine $A$ from the measured symmetry-point lifetimes using the golden-rule relation
\begin{equation}
\Gamma_{1}
=
2\pi uJ(E)\coth\!\left(\frac{\beta E}{2}\right).
\end{equation}
The reported lifetimes of
\SIrange{0.44}{3.2}{\micro\second} at
$\Delta_{0}/h=\SIrange{6.2}{7.34}{\giga\hertz}$
~\cite{Lisenfeld2016,PhysRevB.92.035442}
correspond to values of the dimensionless coupling
\begin{equation}
\alpha\equiv
\frac{J(E^{*})}{k_{B}T/\hbar},
\label{eq:sm:alpha}
\end{equation}
defined in the scaled frequency units of Sec.~\ref{Units+Conv}, between
$4.8\times10^{-8}$ and $5.8\times10^{-7}$ at the operating point.
\SMsubsection{Estimator bias}
The observable that matters is not $\delta p_{e}$ but the bias in $T$ returned by the
estimator of Sec.~\ref{sec:est}. We therefore evaluate Eq.~(\ref{eq:sm:key}) at each point of
the \SIrange{1.3}{2.69}{\giga\hertz} sweep and fit the result with the same
$S=a+b\,p_{e}$ model. Table~\ref{tab:sm:mfgs} reports the outcome as a function of the
bath bandwidth.

Even at the widest band and the most strongly coupled defect, the bias is $68\times$ below the tracking residual; at the weakest coupling it is $820$ times below the \SI{0.082}{\milli\kelvin} residual. The same code reproduces the values of the earlier \SIrange{0.5}{1.6}{\giga\hertz} table to within a factor of 1.5.
\SMsubsection{Numerical validation}
Equation~(\ref{eq:sm:key}) is checked against an independent, non-perturbative
calculation: exact diagonalisation of $H$ for a discretised bath, followed by construction of the exact thermal state and partial trace over the bath. Because the physical $\alpha$ is far too small to resolve against double precision, the coupling is amplified by a factor $s$ and the quadratic scaling is verified. Two truncation errors must be
controlled.

\paragraph{Fock truncation.} The thermal occupation $n(\omega)=(e^{\omega}-1)^{-1}$ reaches $19.5$ at $\omega/\omega_{\rm ref}=0.05$, so an infrared mode needs $\sim\!120$ Fock states. Restricting the exact diagonalisation to $\omega/\omega_{\rm ref}\in[1,8]$, where $n\le0.38$, the ratio of the exact result to Eq.~(\ref{eq:sm:key}) evaluated on the \emph{same} discrete modes is $0.99716$, $0.99776$, $0.99792$, $0.99797$ for Fock cutoffs $M=4,5,6,7$. The second-order expression is thus correct to \SI{0.2}{\percent}.

\paragraph{Mode discretisation.} A coarse logarithmic mode grid overestimates the
continuum. With the same $\omega$ band, the ratio of the discrete-mode evaluation of Eq.~(\ref{eq:sm:key}) to the continuum quadrature converges from $1.42$ at $N = 5$ to $1.0000$ at $N = 2000$; all quoted numbers use the continuum quadrature.

Finally, the bias is carried by modes resonant with the splitting, not by the infrared tail: modes with $\omega>2\,\omega_{\rm ref}$ carry \SI{97.9}{\percent} of the effect. This is the same statement as the $u$ weighting in
Eq.~(\ref{eq:sm:key}), seen from the bath side.
\SMsubsection{The band cutoff is a modelling choice}
The bias in Table~\ref{tab:sm:mfgs} grows with the upper band edge, indicating that the choice of cutoff becomes increasingly important. The numerical ratios for successive factor-of-two widenings are $8.7$, $8.6$ and $6.2$, corresponding to effective exponents of $3.3$, $3.0$ and $2.7$, which approach the expected quadratic scaling from above. Indeed, for $\omega\gg E,k_BT/\hbar$,
\begin{equation}
g(E-\omega)-g(-E-\omega)\simeq \frac{2\beta E}{\omega^{2}},
\end{equation}
so with $J(\omega)=A\omega^3$ the integrand scales as
$2A\beta E\omega$, giving
\begin{equation}
I(E)-I(-E)\propto A\beta E\Lambda^2.
\end{equation}
Thus, the residual bias scales as $\Lambda^2$ rather than $\Lambda^4$. Extending $J(\omega)=A\omega^3$ with the above $A$ to the Debye frequency is unphysical. The resulting Debye–Waller exponent reaches $\sim\!130$, driving $\Delta_0$ to zero. The resolution is standard tunneling-model physics: the high-frequency phonons are already absorbed into the \emph{measured} $\Delta_{0}$ while the hyperbola fit reabsorbs them. We confirmed this directly: widening the bath band in the exact diagonalization moved the recovered $\Delta_{0}$ from $-\SI{19}{\percent}$ to $-\SI{29}{\percent}$. In contrast, the recovered temperature barely moved, because the fit absorbs a dressed $\Delta_{0}$ by construction. What Table~\ref{tab:sm:mfgs} reports is the residual, non-absorbable distortion of the shape of $p_{e}(E)$.

The Debye-frequency check is consistent with this quadratic scaling: scaling the \SI{13.3}{\giga\hertz} entry of Table~\ref{tab:sm:mfgs} by $(8900/13.3)^2$ gives $\sim\SI{0.5}{\kelvin}$, which already lies outside perturbation theory. This is the one place where the analysis rests on a modeling assumption instead of a bound.

There is a second reason to be careful with the cutoff. The relaxation data used to fix $A$ are measured at \SIrange{6}{13}{\giga\hertz}, where the phonon spectrum of the junction is demonstrably not smooth: the expected $\Gamma_{1}\propto\Delta_{0}^{2}/E^{2}$ scaling is obscured by structure in the noise spectral density, and three of five defects share a common relaxation maximum near \SI{7.4}{\giga\hertz}, consistent with coupling to the same confined phonon mode~\cite{Lisenfeld2016}. That interpretation turns on the junction being comparable to a phonon wavelength, which at \SI{7.4}{\giga\hertz} it is ($\lambda\approx\SI{0.9}{\micro\metre}$ for longitudinal phonons in Al, against a $\sim\SI{1}{\micro\metre}$ junction). Over the \SIrange{1.3}{2.69}{\giga\hertz} sweep the wavelength is \SIrange{4.9}{2.4}{\micro\metre}, two to five times the junction, so the continuum $J(\omega)=A\omega^{3}$ is being extrapolated from a band where it visibly fails into a band where the junction cannot confine a mode at all, or where phononic environment modification strongly alters the local acoustic density of states~\cite{ChenPhononEngineering}. A defect-specific measurement of the phonon spectral density in the sub-\SI{2}{\giga\hertz} band would remove both assumptions at once, and is named in the main text as the second quantity an experiment must fix.
\begin{table}[t]
\caption{\label{tab:sm:mfgs}Bias in the recovered temperature from the mean-force
correction, after fitting the full sweep with the three-parameter estimator, for the
strongest and weakest coupling consistent with the measured symmetry-point lifetimes. The systematic floor for comparison is \SI{0.082}{\milli\kelvin}.}
\begin{ruledtabular}
\footnotesize
\setlength{\tabcolsep}{3pt}
\begin{tabular}{@{}lcc@{}}
Bath band & \multicolumn{2}{c}{$|\delta T|$ (mK)}\\
\cline{2-3}
$\omega/2\pi<$ & strongest ($\alpha=5.8\times10^{-7}$) & weakest ($\alpha=4.8\times10^{-8}$)\\
\colrule
\SI{1.7}{\giga\hertz} & $2.6\times10^{-6}$ & $2.2\times10^{-7}$ \\
\SI{3.3}{\giga\hertz} & $2.3\times10^{-5}$ & $1.9\times10^{-6}$ \\
\SI{6.7}{\giga\hertz} & $2.0\times10^{-4}$ & $1.6\times10^{-5}$ \\
\SI{13.3}{\giga\hertz} & $1.2\times10^{-3}$ & $1.0\times10^{-4}$ \\
\end{tabular}
\end{ruledtabular}
\end{table}
\SMsection{Symmetry checks}
\label{sec:symmetry}

Figure~\ref{fig:4} shows the four-level TLS--qubit spectrum and the
strain-dependent quantities used throughout the analysis. The three
symmetry statements of the main text are verified against exact
diagonalisation:
(i) the unitary equivalence
\begin{equation}
H_{\rm TLS}(-\varepsilon)
=
\tau_x H_{\rm TLS}(\varepsilon)\tau_x,
\end{equation}
which gives an even spectrum in $\varepsilon$; (ii) at
$\varepsilon=0$,
\begin{equation}
\left\{H_{\rm TLS},\tau_z\right\}=0,
\end{equation}
and the diagonal elements of $\tau_z$ in the energy eigenbasis vanish which forbids pure dephasing by symmetry; and (iii) the avoided-crossing
gap in the four-level TLS--qubit spectrum is
\begin{equation}
\Delta_{\rm gap}=2g_0\sqrt{u},
\qquad
u=\left(\frac{\Delta_0}{E}\right)^2,
\end{equation}
which agrees with the exact numerical spectrum at the tested strain
points (see Fig.~\ref{fig:4}(b)).

The same parameter $u$ determines the transverse TLS--qubit coupling, $g=g_0\sqrt{u}$,
while the longitudinal contribution scales as $1-u$. Thus, near the hyperbola minimum, $u\simeq1$ and both the transverse coupling and the mean-force temperature bias are maximal, whereas away from the minimum the transverse coupling and associated bias decrease and the longitudinal contribution increases. Figure~\ref{fig:4}(c) shows the corresponding TLS and hybridized frequencies
\begin{equation}
\omega_{\pm}
=
\frac{E/\hbar+\omega_q}{2}
\pm
\frac{1}{2}
\sqrt{
\left(E/\hbar-\omega_q\right)^2+4g_0^2u
},
\label{eq:sm:hybridized_frequencies}
\end{equation}
together with the $E/\hbar$ and $\omega\rightarrow0$ channels relevant to the relaxation, dephasing, and mean-force analyses.

\end{document}